\documentstyle[12pt,fleqn]{article}
\renewcommand{\baselinestretch}{1.2}

\def\fnote#1#2{\begingroup\def\thefootnote{#1}\footnote{#2}\endgroup}

\newcommand{\bea}{\begin{eqnarray}}
\newcommand{\beq}{\begin{equation}}
\newcommand{\eea}{\end{eqnarray}}
\newcommand{\eeq}{\end{equation}}
\newcommand{\nnu}{\nonumber}
\newcommand{\di}{\mbox{d}}
\newcommand{\spav}[1]{\parbox{1mm}{\vspace*{#1}}}

\newcommand{\bx}{\nabla^2}

\newcommand{\bbox}{\bar{\raisebox{-.4ex}{$\bx$}}}

\begin{document}

\begin{titlepage}
\begin{flushright}
CERN-TH.6904/93/Rev.
\end{flushright}
\spav{.1cm}
\begin{center}
{\Large\bf Planckian Energy Scattering} \\
{\Large \bf and Surface Terms in the Gravitational Action}\\
\spav{1.3cm}\\
{ M. Fabbrichesi}\\
{\em CERN, Theory Division}\\
{\em CH-1211 Geneva 23, Switzerland}\\
\spav{.5cm}\\
{ R. Pettorino}\\
{\em Dipartimento di Scienze Fisiche, Universit\`{a} di Napoli}\\
{\em I-80125 Naples, Italy }\\
\spav{.5cm}\\
{ G. Veneziano}\\
{\em CERN, Theory Division}\\
{\em CH-1211 Geneva 23, Switzerland}\\
\spav{.5cm}\\
{ G.A. Vilkovisky}\\
{\em Lebedev Physical Institute and Research Center in Physics }\\
{\em Leninsky Prospect 53, Moscow 117924, Russia}\\
\end{center}
\spav{.5cm}
\begin{center}
{\sc Abstract}
\end{center}
{\small
We propose a new approach to four-dimensional Planckian-energy
 scattering in which the phase of the $\cal S$-matrix is
written---to leading
order in $\hbar$  and
to all orders in $R/b = Gs/J$---in
terms of the surface term of the gravitational action and of a
 boundary term for the
colliding quanta.   The proposal is checked at
leading order in $R/b$ and also
against  some already known examples of scattering in
 strong external
gravitational fields.}

\spav{1.5cm}\\
CERN-TH.6904/93/Rev.\\
November 1993

\end{titlepage}

\newpage

\setcounter{footnote}{0}
\setcounter{page}{1}

\section{Introduction and Outline}

\qquad Planckian-energy collisions represent  a problem ideally
located  between the interests of particle physics and those
of general relativity. At those energies,
 large gravitational fields are both generated
 and felt by the colliding particles, which can then be
 used as sources and probes for classical and quantum gravity
effects.

By changing the impact parameter $b$ of the collision
 (or its total angular momentum, $J=bE$) one can explore large,
intermediate or even short distances, thus progressively increasing
the sensitivity to quantum effects.

 This is particularly noticeable when
the problem is studied within the context of string theory.
String theory possesses a fundamental length parameter of
 its own
\beq
\lambda_s \equiv \sqrt {\alpha ' \hbar} \, .
\eeq
 While General Relativity  expectations are recovered
at large
 distances~\cite{(1)}, substantial quantum-string modifications
occur~\cite{(2)} at
 $b<\lambda_s$ provided  $ R \equiv GE < \lambda_s$.

On the other hand, for $R >\lambda_s$, string-size effects
 appear to be negligible~\cite{(1)}. In this paper we shall assume to be
working
in the latter regime, so that one is left with just two
 dimensionless parameters characterizing the collision:
\beq
\alpha_G \equiv Gs/\hbar \quad \mbox{and}\quad  Gs/J = 4 \, R/b \, .
\eeq

By definition of Planckian-energy scattering, the  parameter
$\alpha_G$, the gravitational equivalent of the fine-structure
constant, will  always be taken to be large. This condition ensures
that the scattering amplitude has a large phase, hence that the
process is semiclassical. We note  that this condition also
 implies that $R$ is much larger than
both
 the Compton wavelength of the colliding particles, $\lambda_c =
\hbar/E$,
 and  the Planck length, $l_P \equiv \sqrt{G\hbar}$.

By contrast, we can still vary at will  the second, crucial
parameter, $R/b$. While  the process is always
  semiclassical, very different physics
 is expected to emerge depending on the value of  $R/b$.

The simplest case occurs at $R/b \ll 1$. Here a leading eikonal
result
 has been obtained by a variety of methods~\cite{(1),(3),(4),(5),Bellini}.
 We note  in particular, for
 later comparison, the approach of Verlinde and Verlinde~\cite{(5)}, where
 the problem is reduced to  estimating a ``topological''
action,  that is, a surface term coming from a reduced two-dimensional
action.

Unfortunately, the most interesting semiclassical phenomena
 that should originate from the collision---such as
 gravitational collapse, black-hole formation and  Hawking
 evaporation---are
 only expected to occur at values of $R/b$ of ${\cal O}(1)$  or less.
 This can be seen in various ways, either within an $\cal S$-matrix
 approach (which gives~\cite{(6),Bellini} higher-order corrections as a power
 series in $R/b$) or from the General Relativity point of view
 (numerical studies of gravitational collapse of rotating
systems~\cite{(7)},
collisions of black holes at high energy and small
 impact parameters~\cite{old d,(8)}, etc.).

It has been pointed out~\cite{(9)} that the approach of
ref.~\cite{(5)} cannot
 be extended in a straightforward way so as to be able
 to cope with this interesting regime.
So far, the most promising way to tackle this difficult problem is,
in our opinion, the one of ref.~\cite{(6)}, where, following work by
Lipatov~\cite{(10)},
 one tries to describe the whole series of corrections in $R/b$ in terms of
 the classical solutions of an effective two-dimensional action
(in the transverse coordinates).

The main criticism that one can raise to such an approach
 is its insistence in separating transverse from longitudinal
 coordinates even in the regime of large angle scattering, which
 necessarily has to precede that of collapse. Besides, the approach
 is technically quite complicated and, so far, only the next-to-leading
 correction has been computed and found to agree with previous
 direct calculations.

In this paper we  propose a completely new approach,
 which remains genuinely four-dimensional at all stages and
 yet yields a ``topological'' result for the ${\cal S}$-matrix. This
approach  is built upon two main ingredients:
\begin{itemize}
\item
 The correct treatment of the gravitational action
through the inclusion of an appropriate surface term;
\item
 A first quantized  path integral  approach to
quantum-field
 theory developed by Fradkin some thirty years ago.
\end{itemize}

It makes it possible to express the semiclassical phase,
 to all orders in $R/b$, just  using surface terms. These come
partly from the surface term of the gravitational action
 and partly from a boundary term connected to the external particles.

The outline of the paper is as follows: In section~2 we
 recall  some basic facts about the necessity
 and   form of the surface term to be added to the usual
 Einstein-Hilbert action.  We also present some general results on how
 to express such a surface integral for asymptotically flat
space-times in terms of Bondi masses. Details of the calculations
pertaining to this section
 can be found in the Appendix.
In section~3 we use Fradkin's approach to
quantum field theory in order  to give convenient
expressions for the full and amputated
 scalar propagators in an external gravitational field. The
 eikonal approximation is recovered in the relevant limit, but
we shall not be restricted
 to it in the following.
In section 4 we   discuss scattering in an external gravitational
 field, easily recovering known results.  We  deal,
 in particular, with the case of an external shock-wave and with
that of  Schwarzschild's metric.
In section 5 we  combine our previous results to obtain a simple
 expression for the $\cal S$-matrix of two colliding particles, first in
 the case of elastic scattering and then with the inclusion of
 gravitational bremsstrahlung. In both cases the process is completely
 determined by surface terms. For the case of elastic scattering we
 see the external metric approximation emerging at leading order
 (and failing beyond) and we recover in an elegant way the leading eikonal
result.
Section 6 contains some remarks concerning the possibility of
extending the method to the case in which graviton loops are included.

\section{The gravitational action and
 the $\cal S$-matrix of massless
 particles}
\setcounter{equation}{0}

\qquad Two related facts distinguish the theory of gravitational
interaction in an asymptotically flat space-time. First, the Einstein-Hilbert
Lagrangian is linearly homogeneous in the metric, whereas the usual
field-theoretic
Lagrangians are essentially quadratic in the field. Second, the energy of
gravitating fields is given by a surface contribution at the asymptotically
flat infinity, whereas, in the non-gravitational field theories, it is the
integral of a volume density. These two properties have not
received proper attention in quantum theory. The reason is probably that, in
perturbation  theory, the gravitational interaction is really not too
different from the  others. It seems, however, that any reasonable
attempt at  non-perturbative  quantum
gravity should rely on the above basic facts. No such attempt will be
undertaken in the present work but we shall argue that,
even in the regime where the gravitational field is treated classically,
the use of the above  features of  gravity  can be fruitful.

\subsection{The gravitational action}

\qquad We start by recalling that the Einstein equations
with   asymptotically flat boundary conditions do not follow from the
Einstein-Hilbert action\fnote{\dag}{We use the signature $(-, +, +, +)$,
the conventions ${\cal R}_{\nu \mu \alpha}^{~~~~~\beta} =
\partial_{\nu}\Gamma_{\mu
\alpha}^{\beta}\, - ...\,$, ${\cal R}_{\nu \mu} = {\cal R}_{\nu \beta
\mu}^ {~~~~~\beta}$, ${\cal R} = g^{\mu \nu} {\cal R}_{\mu \nu}\, ,$ and, in
what
follows,  put $c\,=\, \hbar\,=\, 1 \, .$}
\beq
S_E \,=\,
- \frac{c^{3}}{16 \pi G} \int \di ^{4} x \,\sqrt{-g} \,{\cal R} \, .
\label{2.1}
\eeq
 Indeed, by making a variation of $g_{\mu \nu}$ in a compact domain
$\Omega$,
one finds
\bea
\lefteqn{
\delta \int_{\Omega} \di^{4}x \sqrt{-g}\, {\cal R} \, =}  \nnu \\
& &
 \int_{\Omega}\di^{4} x
\sqrt{-g}
\Biggl[ \left( g^{\mu \nu}\bx \delta g_{\mu \nu}\,- \nabla^{\mu} \nabla^{\nu}
\delta g_{\mu \nu} \right)
 -\left( {\cal R}^{\mu \nu} - \frac{1}{2}\, g^{\mu \nu} {\cal R}
\right) \delta g_{\mu \nu} \Biggr] \, , \label{2.2}
\eea
where $\bx \,=\, g^{\alpha \beta} \nabla_{\alpha} \, \nabla_{\beta}$.
The total derivative  in (\ref{2.2}) can be written in terms of
contributions
from the boundary $\partial \, \Omega$, but when the spatial boundary is
pushed to infinity with the appropriate asymptotic behaviour
 of the
metric, these contributions
do not  vanish. For the Einstein equations to extremize the action, one
should  add to the Einstein-Hilbert action some functional of the
metric  to compensate the contribution of the total derivative in
(\ref{2.2}). We  denote this functional by $S_K[g].$ The total action is
thus of the form
\beq
 S \,=\, - \frac{1}{16 \pi G} \int_{\Omega} \di^{4} x \sqrt{- g} \,{\cal R} +\,
 S_K[g] \, + \, S_{Source}\, , \label{2.3}
\eeq
where we have included possible matter sources of the gravitational field.
The only condition defining $S_K[g]$ is that
\beq
\delta S_K[g]\,=\, \frac{1}{16 \pi G} \int_{\Omega}\di^{4}x \sqrt{-g}
\Bigl( g^{\mu
\nu} \bx \delta g_{\mu \nu}\,-\, \nabla^{\mu} \nabla^{\nu} \delta
g_{\mu \nu} \Bigr) \, , \label{2.4}
\eeq
which ensures that the variational equations of the action (\ref{2.3})
are of the form
\bea
{\cal R}^{\mu \nu} - \frac{1}{2}\, g^{\mu \nu} {\cal R} & = & -8 \pi G\,
{\cal T}^{\mu \nu}\, , \nnu \\
{\cal T}^{\mu \nu} & =  &\frac{2}{\sqrt{- g}} \, \frac{\delta S_{Source}}
{\delta g_{\mu \nu}}\, . \label{2.5}
\eea
Since the action is defined up to an additive constant, we impose
also
the normalization condition that $S_K[g]$ vanishes when the metric is flat.

The derivatives acting on $\delta g_{\mu \nu}$ in (\ref{2.4}) can be
decomposed into
normal and tangential components with respect to the boundary $\partial\,
\Omega$.  Terms with tangential derivatives can be added and subtracted at
will because the action
(\ref{2.3}) is varied under the condition
\beq
\delta g_{\mu \nu}\left|_{~\atop\partial \Omega}\,=\, 0\right. \, .
\eeq
For this reason, there are many functionals $S_K[g]$ satisfying the above
requirements,   differing from each other by tangential derivative terms.
Addition
of such terms affects neither the variational principle nor the
value of the action (\ref{2.3}) computed with an asymptotically flat
metric. The action (\ref{2.3}) with $\Omega$ extended to the whole of
space-time is thus unique.

By using the freedom in tangential derivative terms, the functional $S_K[g]$
can be put in an elegant form \cite{Hawking}, where it is the trace $K$ of the
second fundamental form $K_{ij}$ on $\partial \, \Omega$ integrated over the
boundary. However, we will not be able to use this expression, at least
directly,  because, as
discussed below, we  have to deal with null boundaries. We use instead
the
simplest expression for $S_K[g]$, the one that  can be obtained as
follows.

Let the boundary $\partial \, \Omega$ be given by the equation
\beq
 \partial \, \Omega \, : \, \tau (x)\,=\, 0 \, ,
\eeq
where $\tau (x)$ is a piecewise smooth function, and
let the sign of $\tau (x)$
be chosen so that the direction of growing $\tau$ leads out of $\Omega$.
Then the Gauss theorem can be put in the form:
\beq
\int_{\Omega}\di^{4} x \sqrt{-g}\: \nabla_{\mu} f^{\mu}\,=\, \int \di^{4}x
\sqrt{-g}\: \delta \left(\tau (x)\right) f^{\mu} \nabla_{\mu} \tau\, ,
\eeq
which is also valid for null boundaries, $ \left(\nabla \tau
(x)\right)^{2}\,=\, 0$. Accordingly, eq. (\ref{2.4}) takes the form:
\beq
\delta S_K[g] \,=\, \frac{1}{16 \pi G} \int \di^{4} x \sqrt{-g}\: \delta\!
\left(\tau (x) \right) \, \nabla_{\mu} \tau \left( g^{\alpha \beta}
\nabla^{\mu} \delta g_{\alpha \beta}\, - \, g^{\mu \alpha} \nabla^{\beta}
\delta g_{\alpha \beta}\right) \, .
\eeq
Since the metric enters $S_K$ only near the boundary, and the boundary
will
be carried to the domain where  space-time is asymptotically flat,
we can introduce an auxiliary flat metric $\tilde{g}_{\mu \nu}$ in this
domain and expand $S_K$ as
\beq
S_K\,=\, S_K[\tilde{g} + h]\,=\, A[\tilde{g}]\,+\, B [\tilde{g}, h]\, +\,
{\cal O}[h^{2}]\, ,
\eeq
where $A$ is the zeroth-order term, $B$  the first-order term, and
${\cal O}[h^{2}]\,$ higher-order terms. With the  above normalization,
\beq
A\,[\tilde{g}]\,=\,0\, ,
\eeq
and the ${\cal O}[h^{2}]\,$ terms fall off at infinity too rapidly to
give a finite contribution. The
result is that, whatever $S_K$ is, it is given by a linear term of its
deviation from the flat space-time value. But then we already know it, it
is:
 \beq
S_K \, = \, \frac{1}{16 \pi G}
\int_{\!\!{~\atop{{asymptotic}\atop{domain}}}}\!\!\!
\di ^{4} x \sqrt{- \tilde{g}}\, \delta (\tau(x))\,  \nabla_{\mu} \tau
\left(\tilde{g} ^{\alpha \beta} \tilde{\nabla}^
{\mu} h_{\alpha \beta} \, - \, \tilde{g}^{\mu \alpha}
\tilde{\nabla}^{\beta}
h_{\alpha \beta} \right)\, , \label{2.13}
\eeq
where the quantities and operators with tilde refer to the flat metric
$\tilde{g}_{\mu \nu}$, and
\beq
h_{\mu \nu}\,=\, g_{\mu \nu}\,-\, \tilde{g}_{\mu \nu} \, .\label{wf}
\eeq
The components of $g_{\mu \nu}$ are calculated in a
chart
covering the asymptotic domain, and the components of $\tilde{g}_{\mu \nu}$
are their flat space-time limits.

The reason for the
surface term (\ref{2.13}) to be non vanishing lies in the fact
 that the expansion of the
Einstein-Hilbert action in the wave field (\ref{wf}) starts with a linear
term. Hence, the surface term to be subtracted is also linear in the field
and, finally, the field itself is not decreasing sufficiently
 fast at large distance
for giving a vanishing contribution.

 The boundary contribution to the gravitational action has
been discussed
in the literature mainly in the context of Euclidean gravity. We need to
calculate it for a Lorentzian asymptotically flat space-time. A major
requisite of this calculation is the determination of the boundary of an
asymptotically flat space-time. If one thinks of  space-time as of a
``cylinder'' bounded by time-like and space-like hypersurfaces, then,
at the space-like
portions, the metric will not be asymptotically flat at all.  The boundary
 of an  asymptotically flat space-time emerges after the Penrose
conformal transformation, bringing infinity to a finite
distance~\cite{Penrose}.
 It then consists of the spatial infinity $(I^\circ)$, where the
space-like geodesics begin and end, the past and future time-like infinities
($I^{-}$ and $I^{+}$),
where the time-like geodesics begin and end, and the past and future null
infinities (${\cal I}^{-}$ and ${\cal I}^{+}$), where the null geodesics
begin
and end. In the physical space-time, the $I^{\circ},\, I^{\pm}, \, {\cal I}^
{\pm}$ are understood as infinite limits of affine parameters along the
respective geodesics. The metric is asymptotically flat
at ${\cal I}^{-}\,,\,
{\cal I}^{+}$ and $I^{\circ}$,
but not at $I^{-}$ and $I^{+}$. An
important fact is, however, that $I^{\circ}\,,\, I^{-}$ and $I^{+}$ are
single points, whereas ${\cal I}^{-}$ and $ {\cal I}^{+}$ are
three-dimensional
(null) hypersurfaces. Since the surface term in the action is an integral
over the boundary, we conclude that it is the integral over
\beq
\partial \Omega\,=\, {\cal I}^{-} \cup {\cal I}^{+} \, .
\eeq
If  space-time has event horizons, the latter also appear as portions
of the boundary. We shall come back to this point when discussing the
$\cal S$-matrix.

More knowledge about the boundary is given by the behavior of the metric in its
 neighborhood.
 The past and future null infinities are treated
similarly. In  the appendix  we shall describe ${\cal I}^{+}$ following
Sachs \cite{Sachs}, and   build the metric near ${\cal I}^{+}$
by considering
the congruence $ u(x)=\mbox{const}, \, (\nabla u)^{2}\equiv 0$, of light
rays  reaching ${\cal I}^{+}$. A similar procedure can be followed
at ${\cal I}^{-}$. In the same appendix, we then  calculate the boundary
term in the action and show that it can be written in the form
\beq
S_K \, = \,- \frac{1}{2} \left( \int_{- \infty}^{\infty} \di u\, M_{+} (u) -
\int_{- \infty}^{\infty} \di v\, M_{-}(v) \right)\, , \label{2.16}
\eeq
where $M_{+}(u)$ and $M_{-}(v)$ are the Bondi masses at ${\cal I}^{+}$ and
${\cal I}^{-}$, the precise definition of which
is given in the  appendix, and the
retarded and advanced time are normalized by the  conditions
(\ref{A.16}) and (\ref{A.30}).

\subsection{The ${\cal S}$-matrix of massless fields}

\qquad The ${\cal S}$-matrix of the gravitational field coupled to a set of
matter fields $\psi$ is given by the functional integral
\beq
{\cal S}\,=\, \int \di [g_{\mu \nu}, \psi ]\, e^{i S} \, , \label{2.19}
\eeq
where $S$ is the full action (\ref{2.3}), and the integration is carried
out over all fields interpolating between the asymptotic fields with
fixed operator data. For
massless conformal invariant fields, these
data  are at  ${\cal I}^{\pm}$.
It is assumed that the measure in (\ref{2.19}) includes all gauge-fixing,
ghost and $ \delta(0)$ contributions.

To leading order in  $\hbar$,  the classical path dominates the
functional
integral, and the ${\cal S}$-matrix takes the form
\beq
{\cal S}_{tree}=\, \exp \Bigl( i S\mid_{sol'n} \Bigr) \, ,\label{ST}
\eeq
where the action is to be calculated on the solution of its variational
equations with the appropriate boundary conditions.
This approximation amounts to a resummation
 of the  Feynman diagrams illustrated in fig. 1, as discussed further in
section 5.

 Consider first the
case of pure gravity. By virtue of the field equations (\ref{2.5}), the
volume density of the Lagrangian in (\ref{2.3}) vanishes, and the action
reduces to the  surface term only:
\beq
{\cal S}_{tree} =\, \exp \Bigl( i\, S_K[g] \mid_{sol'n} \Bigr) \, . \label{SS}
\eeq
 Then, by using the result
(\ref{2.16}),
we  obtain:
\beq
{\cal S}_{tree}\,= \exp \left[- \frac{i}{2} \left(\int_{- \infty}^{\infty}
\di u\, M_{+} (u) - \int_{- \infty}^{\infty} \di v\, M_{-} (v)\right)
\right] \, ,\label{2.20}
\eeq
which relates the $\cal S$-matrix elements directly to the Bondi masses of the
classical gravitational radiation.

The vanishing of the Einstein-Hilbert action in (\ref{ST}) is, of course,
based on its being homogeneous in the metric. However, in general,
 homogeneity alone is not  sufficient for the action to vanish.
With $S(\varphi)$ homogeneous of degree $n$ in $\varphi$, and
$\varphi$ infinite-dimensional (e.g. a field), one can only
 conclude that:
\beq
n  \,S(\varphi) \,=\,
\int \di x\, \varphi (x)\,{\frac{\delta \,
 S(\varphi)}{\delta \,\varphi}}
\,+\,
\mbox{a surface term}\, . \label{euler}
 \eeq

The key point
here is that the equations of motion are obtained from variations
 of $\varphi$ which vanish at
the boundary, while the variation of $\varphi$ needed in the homogeneity
(Euler) equation for $S$, being global, does not satisfy this property.
With the corrected action (\ref{2.3}) for pure gravity, one arrives at
the result (\ref{SS}), that is to the identification of the
surface term in (\ref{euler}) with $S_K$. We stress once more that
the reason for the surface term
to be non-vanishing in this case lies in the fact that, for gravity,
 the surface term is linear
in a field which is
not decreasing sufficiently fast at large distance.

If one considers, instead of the ${\cal S}$-matrix,  the generating
functional for Green's functions
\beq
Z[J]\,=\, \int \di [g_{\mu \nu},\, \psi] \exp \left(i S \,+\, i \int \di^{4} x
\sqrt{-g} g^{\mu \nu} J_{\mu \nu} \right) \, , \label{2.21}
\eeq
and chooses the field variables as $g_{\mu \nu}$ or $ \sqrt{-g} g^{\mu
\nu}$,
the result will again be of the form (\ref{2.20}), with the only
difference that the
classical equations will be modified by the presence of an external source.
The total volume density of the Lagrangian in (\ref{2.21}) is again
linearly  homogeneous in $g_{\mu \nu}$, and no additional surface terms
appear. With other choices of the field variables, the result (\ref{2.20})
will hold only on shell.

The result (\ref{2.20}) remains valid if the gravitational field is
coupled to non-self-interacting conformal invariant matter fields.
Indeed, the quadratic actions  of these fields are reduced to surface terms
by their own
equations of motion, and, unlike the gravitational surface term, they
  vanish provided  one chooses the solutions of the
source-free equations that vanish at infinity and have finite energies.
Since the energy-momentum tensor of conformal invariant fields is traceless,
the total action is again of the form (\ref{SS}) although the solution
to be used in (\ref{SS}) is, of course, different.
The absence of self interactions combined with conformal invariance of
matter fields is a sufficient but not necessary condition for the action to
reduce to the surface term $S_{K}[g]$.  For example, the action
\beq
- \frac{1}{2} \int \di ^{4}x \sqrt{-g}\, g^{\mu \nu} \nabla_{\mu} \psi
\nabla_{\nu}
\psi \,
\eeq
of a scalar field $\psi$ is not conformal invariant but is linearly
homogeneous
in $g_{\mu \nu}$, so that (\ref{SS}) again holds.

 A necessary condition
follows from the form of expression (\ref{2.20})
\fnote{\dag}{One of the authors (G.A.V.) is grateful to Don
Page for discussing this point.}. Note that $M_{+}(u)$ has finite
limits at both $u= - \infty$ and
$u= + \infty$,
and similarly $M_{-}(v)$. The limits $M_{+}(- \infty)$ and $M_{-}(+
\infty)$ are
both equal to the $ADM$ mass $M_{0}$, so that $M_{0}$ cancels in the
difference.  If  matter consists only of massless radiation that
comes in through  ${\cal I}^{-}$ and goes out through ${\cal I}^{+}$,
one has $M_{-}( - \infty)=
0$ and $M_{+}(+ \infty)=0.$
Generally, for the difference of the integrals in (\ref{2.20}) to be
finite, one  should at least have
\beq
M_{-}(- \infty)\,=\, M_{+}(+ \infty) \, ,
\eeq
which means that the energy carried by time-like sources is conserved
separately.
If this condition does not hold, there should be a non-vanishing volume
density
of the Lagrangian to secure the finiteness of the total action.

A promising feature of writing the action through the
Bondi masses, is that it easily allows for the introduction of the
machinery and interpretations of classical gravity theory,
thus making possible the use of
 solutions~\cite{old d,(7),(8)} that go far beyond flat-space perturbation
theory.  The ${\cal S}$-matrix can then be constructed and used, hopefully, for
studying the semiclassical phenomena discussed in the Introduction.
In  particular, if the relevant classical solution has
event horizons, then not all the energy entering through ${\cal I}^{-}$ will
appear at ${\cal I}^{+}$; a portion of it will fall into the black
hole. Since the ${\cal S}$-matrix is calculated only between the states
defined at ${\cal I}^{-}$ and  ${\cal I}^{+}$, this loss should be felt
in (\ref{2.20}). Alternatively, one may  include the horizon as a
portion of the boundary, which means adding the states
defined at the horizon.

\section{Scalar Propagators in Fradkin's Approach}
\setcounter{equation}{0}

\qquad Fradkin~\cite{Fradkin} has pioneered an approach to
quantum field theory,
 in which  integration over a quantum field is avoided
 by writing the propagators as functional integrals of the
first quantized theory. Here we use his formulation
 in the case of
the  propagator for a scalar field in Einstein's theory of
gravity
and in the presence of an electromagnetic field.

\subsection{The Propagator}
\qquad
The Green function (Feynman propagator) for a scalar field
 is defined~\cite{Fock} as
\beq
G(x,y/g) = \langle y | {\cal H}^{-1} | x \rangle = i \int_0^\infty
\di \nu  \langle y | \exp \left[ -i ({\cal H} - i \epsilon )  \nu \right] |
x
\rangle \, ,  \label{Green1}
\eeq
where, in the case of propagation in an external gravitational field
  $g_{\mu\nu}$,
\beq
{2 \cal H} = -\nabla_\mu g^{\mu\nu} \nabla_\nu + m^2 =
\frac{-1}{\sqrt{-g}} \partial_\mu \sqrt{-g} g^{\mu\nu} \partial_\nu + m^2
\, . \label{H}
\eeq

Equation (\ref{Green1}) admits a representation as
a path integral as~\cite{Feynman}
\beq
G(x,y/g) = i \int_0^\infty
\di \nu  \int_{X(0)=x}^{X(\nu)=y} [\di X^\mu ]\,[\di P_\mu]
\exp \left[ i\int_0^\nu \di \nu  \left( P \cdot \dot{X} - {\cal H}
\right) \right] \, , \label{Green2}
\eeq
where the dependence of $\cal H$ on $P^\mu$ is given by
replacing the covariant derivative $\nabla^\mu \rightarrow  i
p^\mu$. The correctness of this procedure can also be verified by
considering the action of a relativistic point in an external
metric: \beq S = -m\int \di \lambda \sqrt{-g_{\mu\nu}\dot{X}^\mu
\label{action} \dot{X}^\nu} \, .
\eeq
This can also be written as
\beq
S= -\frac{1}{2} \int \di \lambda \left[ e^{-1/2} \left(
-g_{\mu\nu}\dot{X}^\mu
\dot{X}^\nu \right) + e^{1/2} m^2 \right] \label{a1}
\eeq
by introducing the auxiliary einbein $e$. The action (\ref{a1}) has
the advantage of allowing a straightforward massless limit $m \rightarrow
0$.
Eliminating $e$ through its equation of motion:
\beq
e = -g_{\mu\nu}\dot{X}^\mu\dot{X}^\nu /m^2 \label{b}
\eeq
one recovers (3.4).

According  to (3.5), the conjugate momentum is
\beq
P_\mu = e^{-1/2} \dot{X}^\nu g_{\mu\nu} \label{a}
\eeq
so that, upon use of (\ref{a}) and (\ref{b}):
\beq
P_\mu g^{\mu\nu} P_\nu  =  g_{\mu\nu}\dot{X}^\mu
\dot{X}^\nu /e = -m^2 \, .
\eeq

The reparametrization invariance of the action (\ref{action}) forces the
canonical Hamiltonian to vanish so that the full Hamiltonian reduces to
the constraint itself, that is
\beq
{\cal H} = \frac{1}{2} \left( P_\mu g^{\mu\nu} P_\nu  +m^2 \right) \, .
\label{3.9}
\eeq

In the definition (\ref{3.9}) we have neglected  ordering ambiguities, which
give rise to a possible additional term proportional to the scalar
curvature.
The coefficient of such a term can be fixed, for example, by imposing
conformal
invariance for the massless, spin-0 wave equation, to obtain
\beq
2 {\cal H} =   -\nabla_\mu g^{\mu\nu} \nabla_\nu + m^2 -\frac{1}{6}\, {\cal
R}
\label{1/6} \, .
\eeq
However, in this paper, the functional integral (\ref{Green2}) will be used
only
to the lowest-order WKB approximation, where it is unambiguous, and the
problems
 inherent in its definition
will not be discussed further.

 The propagator is
therefore defined by (\ref{Green2}), which yields, after integrating out the
momenta,
\bea
\lefteqn{G(x,y/g) = \int_0^\infty \di \nu  \int [\sqrt{-g(X)} \di
X^\mu ]
\delta^{(4)} \left(x - X(0) \right)  \delta^{(4)} \left( y - X(\nu) \right)}
\nnu \\
& & \times \exp \left[ (i/2) \int_0^\nu \di \xi \left(
g_{\mu\nu}\dot{X}^\mu
\dot{X}^\nu  - m^2 \right) \right] \label{Green3} \, .
\eea
The measure in (\ref{Green3}) and below must be understood as containing the
usual factor $(2\pi \di \nu )^{-2}$ arising in the integration over $[\di P_\mu
]$.

 The Green function (\ref{Green3})  can also
be generalized  in a straightforward fashion
to include an electromagnetic field by the usual
substitution $p_\mu \rightarrow p_\mu -eA_\mu$ to give
\bea
\lefteqn{G(x,y/g,A) = \int_0^\infty
\di \nu  \int [\sqrt{-g(X)} \di X^\mu ]  \delta^{(4)} \left(x - X(0)
\right)  \delta^{(4)} \left( y - X(\nu) \right)} \nnu \\
& & \times \exp \left[ (i/2) \int_0^\nu \di \tau \left(
g_{\mu\nu}\dot{X}^\mu \dot{X}^\nu  +2eA_\mu \dot{X}^\mu - m^2
\right) \right] \label{Green4} \, .
\eea

The propagators (\ref{Green3}) and (\ref{Green4}) are connected
(in the sense that the vacuum diagrams have been
already  divided out), but not amputated.  In order to
properly  define the
 ${\cal S}$-matrix \`{a} la LSZ, we need the on-shell amputated propagator,
  in which the external legs have been removed:
\bea
\lefteqn{G_c(p,p'/g, A) =} \label{def}  \\
& &  \lim_{p^2,p'^2\rightarrow -m^2} (p^2 +m^2) (p'^2
+m^2) \int \di^4 x \: \di^4 y \exp \Bigl[ i p\cdot x -ip'\cdot
y \Bigr] G(x,y/g, A)
\, , \nnu
\eea
where the Fourier transform is taken with respect to asymptotically
Cartesian
coordinates. The  prescription we use for obtaining  $G_c$ directly is the
following~\fnote{\dag}{
This method, based on a result given in~\cite{MF}, was used within a
functional approach
 in a series of papers on scattering amplitudes in QED~\cite{barba}.}.
 The integral over $\nu$ in  (\ref{Green3}) and
(\ref{Green4}) is
 replaced by a limit in which the initial and final $\nu$'s
go, respectively, to minus and plus infinity. Accordingly, the amputated
propagator is given by
\beq
G_c(p,p'/g, A) = \lim_{p^2,p'^2 \rightarrow -m^2}
\lim_{\nu_i \rightarrow -\infty
\atop
\nu_f \rightarrow \infty} \frac{1}{\nu_f-\nu_i}
\int [ \sqrt{-g(X)} \di X^\mu] \exp \Bigl[ i
{\cal A}  \Bigr] \, ,\label{Gc}
\eeq
where
\beq
{\cal A} = p \cdot X(\nu_i) - p' \cdot X(\nu_f) +
\int_{\nu_i}^{\nu_f} \di \tau L(\tau) \label{A}
\eeq
and
\beq
L(\tau) = \frac{1}{2} \left( g_{\mu\nu} \dot{X}^\mu
\dot{X}^\nu + 2eA_\mu \dot{X}^\mu- m^2 \right) \label{L} \, .
\eeq

The prescription (\ref{Gc}) gives the same result as the
definition (\ref{def}), as  can be readily checked in
perturbation theory.

 The form (\ref{L}) of the Lagrangian
corresponds to the gauge in which $e=1$ and $\lambda
=\tau/m$, where $\tau$ is the particle's proper time. Only
affine transformations remain as  invariances of the action
after such a gauge fixing.

 The propagator (\ref{Gc}) is our
starting point. It provides a solution  for the quantum motion of a
scalar particle in terms of a functional integral over trajectories.
Notice that the measure  $[\di X^\mu ]$ in (\ref{Gc})  includes an
integration over the initial ($X^\mu(\nu_i)$) and final
($X^\mu(\nu_f)$) positions.

At Planckian energies the mass term as well as the electromagnetic
interactions can be neglected. Furthermore the action $\cal A$ is large
   and the
stationary phase approximation should be valid to
${\cal O} (\hbar^{-1})$.

Making $\cal A$ stationary not only with respect to  $X^{\mu}(\tau)$
inside the integration region for $\tau$ but also, say, with
respect to the final position gives:
\beq
\ddot{X}^\mu + \Gamma^\mu_{\nu\sigma} (X)
\dot{X}^\nu\dot{X}^\sigma  = 0,\label{geod}
\eeq

\beq
\dot{X}^\nu g_{\mu \nu}(\nu_f) = \dot{X}^\nu(\nu_f) \eta_{\mu \nu} =
p'_{\mu}. \label{geodd}
 \eeq

At this point the only integration left is an ordinary integration over the
initial position, everything else being then fixed by the (null) geodesic.
As a result the bulk term in
(\ref{A})  vanishes and the phase of the Green function (\ref{Gc})
  (phase shift) can be computed in terms of  the initial and final
positions  of the particle as:
\bea
G_c (p,p'/g) & \simeq & \mbox{Lim}
\int \di^4 X_i
\exp i\Bigl[  p \cdot X_i - p' \cdot X_f (X_i)
  \Bigr] \nnu \\
& \simeq &
 \mbox{Lim}
\int \di^4\Sigma
\exp -i\Bigl[  q\cdot\Sigma + P \cdot \Delta (\Sigma )
\  \Bigr] \label{prop} \, ,
\eea
where we have introduced the notations:
\bea
X_i  \equiv  X(\nu_i), & &\; X_f \equiv X(\nu_f), \; \nnu \\
  \Sigma = \left(
X_i + X_f \right)/2, & &\; \Delta = X_f -X_i \, , \nnu \\
 P =  \left( p+p'\right)/2,  & &\; q=p'-p \, ,
\eea
  and we have taken $\sqrt{-g(X_i)} =1$ since
 the asymptotic
 conditions are defined far enough from any gravitation field and
we are using Cartesian coordinates there. Finally, we have
introduced the short-hand notation:
\beq
\mbox{Lim} \equiv \lim_{p^2,p'^2 \rightarrow -m^2}
\lim_{\nu_i \rightarrow -\infty
\atop
\nu_f \rightarrow \infty} \frac{1}{\nu_f-\nu_i} \, .
\eeq

The final expression for the
propagator (\ref{prop}) only depends on $X_i$ and $X_f$: it has
thus been reduced to a ``boundary'' term.
The propagator (\ref{prop}) appears to take
different forms according to which variables are integrated first
 by means
of the saddle-point approximation (i.e. by using the geodesic) and which
are left to the end for a more precise integration method. In
(\ref{prop}) we have written two possible forms for the propagator that we
will use in the following. In the second one, we have
traded the remaining integration over $X_i$ for the one over
$\Sigma$, with the understanding that $\Delta$ should be expressed
in terms of $\Sigma$ and of the external field.

  The final outcome of such
a procedure may depend, in general, on the above separation of integration
variables, the correct choice being dictated by the particular
problem at hand.  In the following, in order to be on the safe side,
 we shall
always imply   that our final results, when expressed as integrals, are to be
trusted only in the saddle-point approximation, where the
order of integration becomes immaterial and the result is unambiguous.

\subsection{Eikonal Approximation}

\qquad As a check of (\ref{Gc}) and of (\ref{prop}), we can compute $G_c$
in the
eikonal limit. This  corresponds to a saddle-point approximation
for the functional integral in which the  classical trajectory
$X^\mu(\tau)$ is computed to the lowest non-trivial order.  In
this case, the Green function (\ref{Gc}) feels the change in phase along
such a trajectory as a function of the external field $g_{\mu\nu}(X)$ or
$A_\mu (X)$.

Let us   consider again the massless case $m=0$
in the absence of electromagnetism. We write
 the metric tensor as
\beq
g_{\mu\nu} = \eta_{\mu\nu} + 2\kappa h_{\mu\nu}
\, ,
 \eeq
where $\kappa^2 = 8\pi G \, .$

Solving for the geodesic motion (\ref{geod}) to lowest order in $\kappa$
yields:
\beq
\dot{X}^\mu (\nu) = \dot{X}^\mu (\nu_i) -
\dot{X}^\nu (\nu_i) \dot{X}^\sigma (\nu_i) \int_{\nu_i }^
{\nu} \di
\tau \: \Gamma ^\mu_{\nu\sigma} (X(\tau))
\eeq
and thus, after use of (\ref{geodd}),
\beq
X^\mu (\nu) =  X^\mu _i  + p^{\mu} (\nu - \nu_i) -
p^{\nu} p^{\sigma}
  \int_{\nu_i} ^{\nu }
\di \tau \int_{\nu_i}^\tau \di \tau' \:
\Gamma ^\mu_{\nu \sigma} (X(\tau')) \, .
\eeq

 We can now insert the above perturbative results
into (\ref{A}) to obtain
 \beq
 {\cal A}  = -q\cdot \Sigma +
  \kappa\int_{\nu_i}^{\nu_f}
 h_{\mu\nu} \left( X(\tau) \right) p^\mu p^\nu \di \tau + {\cal O}(q^2)
+ {\cal O}(\kappa q)
\, . \label{ph}
 \eeq

In the argument of $h_{\mu\nu}$ we may now use
the lowest-order
(straight) trajectory:
 \beq
X^\mu_{0} (\nu) =
\Sigma^{\mu} + p^\mu \left[ \nu -  (\nu_f +\nu_i)/2 \right] \, ,
\eeq
where
\beq
  p^\mu = \frac{\Delta ^{\mu}}{\nu_f - \nu_i} \, .
\eeq

 Inserting these results into (\ref{prop}), the
  eikonal  propagator becomes
 \bea
\lefteqn{G_c(p,p'/h)  =
\lim_{\nu_i \rightarrow -\infty
\atop
\nu_f \rightarrow \infty} \frac{1}{\nu_f-\nu_i}
\int \di ^4 \Sigma}  \nnu \\
& & \times
\exp \left[ -iq \cdot \Sigma +
 i\kappa\int_{\nu_i}^{\nu_f}
 h_{\mu\nu} \left(\Sigma +
 p  \, ( \tau -  \frac{\nu_f +\nu_i}{ 2})
 \right)  p^\mu p^\nu \di \tau \right] \, ,
\label{eik}
 \eea

  The component of $\Sigma$ parallel to $p$ can be easily integrated since
the integrand depends trivially on it:
 \beq
 \int \di^4 \Sigma \exp \left[ -iq\cdot \Sigma \right]  (\dots ) =
p^0  \int_{\nu_i}^{\nu_f}  \di \nu \:
 \exp \left[- i\frac{q \cdot p}{2} \nu \right] \int \di^3 \Sigma
   \: ( \dots ) \, .
\eeq
The integration in $\di \nu$ cancels, to leading order in $q$,
 the factor
$1/(\nu_f -\nu_i)$ in the limit defining (\ref{Gc}) and we are left with
the integration over the  three components
 of $\Sigma$ orthogonal to $p$.

 After a trivial change of variables we finally obtain:
 \beq
 G_c(p,p'/h) \simeq E \int \di^3 b \:
 \exp \left[ -iq \cdot b +
 i\kappa \int_{-\infty}^{+\infty}
 h_{\mu\nu} \left( b+p\tau \right) p^\mu p^\nu \di \tau \right] \, ,
\label{geik}
 \eeq
where  we have introduced
\beq
 b = \frac{\nu_f X_i - \nu_i X_f}{\nu_f - \nu_i}\, ,
\eeq
so that $X^\mu_0 (\nu) = b^\mu + p^\mu \nu$,
and the remaining integration is with respect to the components of $b$
others than the one in the direction of $p$. The above result is
in agreement with the known expression~\cite{KO}.

The electromagnetic case can also be  derived along similar lines
to give the old result~\cite{eikonal}:
\beq
 G_c(p,p'/A) \simeq E \int \di^3 b \:
 \exp \left[ -iq \cdot b +
  i e \int_{-\infty}^{+\infty}
 A_{\mu} \left( b+p\tau \right) p^\mu \di \tau \right] \, .
\eeq

It is perhaps instructive to give a second derivation of the gravitational
eikonal starting directly from (\ref{Gc}) and using at first only
(\ref{geod}) at fixed $X(\nu_i)$, $X(\nu_f)$. A straightforward
calculation
then gives:
\bea
 {\cal A}  &= & -P\cdot \Delta - q \cdot \Sigma
+ {\Delta^{\mu}\Delta^{\nu}\eta_{\mu\nu}\over 2(\nu_f -\nu_i)}   \\
 & &   +
{\Delta^{\mu}\Delta^{\nu} \over  (\nu_f
-\nu_i)^2} \kappa \int_{\nu_i}^{\nu_f} \di \,\tau \, h_{\mu \nu}\,
\left(\Sigma +
 \Delta   \, { \tau -  (\nu_f +\nu_i)/2 \over  \nu_f - \nu_i
} \right)  \, .\nnu
\eea
The integration over $\Delta$ can be done to lowest order in $\kappa$
by standard perturbative techniques. $\Delta$ is replaced in the
interaction part of the action by a differential operator acting on the
remaining integral, which is Gaussian, to obtain:
 \bea
\lefteqn{G_c(p,p'/g)  = \mbox{Lim} \int \di ^4 \Sigma  \:\exp \left( -iq
\cdot \Sigma
\right)} \label{3.33} \\
& & \times  \exp \left[ \frac{-i}{(\nu_f -
\nu_i)^2} \frac{\partial}{\partial P_\mu} \frac{\partial}{\partial P_\nu}
\kappa \int_{\nu_i}^{\nu_f} \di \tau \, h_{\mu\nu} (\Sigma + \dots) \right]
\exp \left[ -i P^2 (\nu_f - \nu_i ) /2 \right] \, . \nnu
\eea
 Terms in (\ref{3.33}) obtained by not acting  on the
exponent can be shown to be sub-leading by powers of $q/P$ or of
$1/P^2(\nu_f-\nu_i)$ and are thus negligible for small angles.
 One is
therefore left with the differential operators   acting only
on the exponent  and one remaining
ordinary integration variable, which can be chosen to be $\Sigma$. The
result (\ref{geik}) is thus easily  recovered.

\section{Scattering in an External Gravitational Field}
\setcounter{equation}{0}

\subsection{General Considerations}

\qquad  Before moving on
to the full scattering matrix of the two-body collision,
it is useful to check our framework in the simpler case of the
scattering in an external gravitational field.

We consider two examples for which an exact solution for the geodesics is
known: the (generalized) Aichelburg-Sexl (AS) shock wave~\cite{AS}
and the Schwarzschild (black hole) metric~\cite{S}.
Planckian scattering in the Schwarzschild metric has been discussed in
perturbation theory in~\cite{sole}.

\subsection{Scattering by a null shock wave}

\quad The AS metric is an exact solution for the gravitational
field produced by a single massless particle. Its generalization
to an instantaneous light pulse of arbitrary energy profile
$\rho (y,z)$ is of the form:
\beq
\di s^2 = - \di U \,\di V + f(y,z) \delta (U)\,\di U^2 + \di y^2
+ \di z^2 \, ,
\eeq
where  $U=t-x$, $V=t+x$ are flat-space null coordinates and the
 function $f$ of the transverse coordinates $y,z$,  the
profile function of the shock wave, is related to the energy profile $\rho$
by the Einstein equations:
\beq
\Delta ~f = - 2 \kappa^2 \rho \, .
\eeq
For the special case of a point particle of energy $E_b$ ($b$ for beam),
this gives
 the well known AS result, for which
 \beq
\rho = E_b ~\delta (U)~ \delta^{(2)} (x_T)
\eeq
where $x_T = (y,z)$, and
\beq
f(x_T) = -\frac{\kappa^2}{2\pi} E_b ~\ln (y^2 + z^2) \, . \label{shock}
 \eeq
The classical trajectory lies in a plane. In a coordinate system in which
this is the $x-y$ plane it is
given, in parametric form, by~\cite{FPV}
 \bea
 U(\nu) & = & 2p^u\nu +
U_0 \nnu \\ V(\nu) & = & f'^2U(\nu) \theta (U)/4 + f(b)\theta (U) + V_0
\nnu \\ x_T(\nu) & = & \left (b +f'U(\nu)\theta (U)/2 \, , \, 0
\right) \, , \label{SW}
 \eea
 where
$V(\nu_i)=V_0$, $U(\nu_i)=U_0$ and $x_T(\nu_i)=b$ are
 the initial conditions.

We take $\nu_i = 0$ and $\nu_f =\nu$ to simplify the notation; the
propagator can be written according to (\ref{prop}) as
\bea
\lefteqn{ G_c (p,p'/AS)  =   }  \nnu \\
& & \lim_{\nu \rightarrow \infty} \frac{1}{\nu} \int
\di^4\Sigma   \exp i \Bigl[ (q^u\Sigma^v + q^v \Sigma^u)/2 -
q_T\Sigma_T  \Bigr.\nnu \\
& &   \Bigl. + (P^u\Delta ^v + P^v \Delta^ u)/2 - P_T\Delta_T  \Bigr] \, ,
\eea
where, from (\ref{SW}), we have:
\bea
\Sigma^v &=& \frac{1}{2} \left[f(b) + \frac{1}{4} f'^2 \left(2p^u\nu +U_0
\right)\right] + V_0 \, , \nnu \\
\Sigma^u &=& p^u\nu + U_0 \, ,\nnu \\
\Sigma_T &=& \left (b+ \frac{1}{4} f' \Bigl[ 2p^u\nu +U_0 \Bigr]\, , \, 0
\right) \, ,
\eea
and we parametrize the momenta as follows:
\bea
P &=& {1\over 2} \left( E+E', \: -E-E'\cos\vartheta, \: -E'\sin
\vartheta, \: 0  \right) \, , \nnu \\
q &=& \left( E'-E, \: E-E'\cos\vartheta, \: -E'\sin \vartheta, \: 0
\right) \, .
\eea

Integration over $V_0$ gives a  $\delta(q^u)$ thus
enforcing
\beq
E=E' (1+\cos\vartheta)/2 \, ,
\eeq
so that
\beq
P^u = 2E\, , \qquad  q^v = 2 P^v = 2(E'-E) \quad \mbox{and} \quad
2 P_T=q_T=-E'\sin\vartheta \, .
\eeq
Accordingly, the phase becomes
\beq
- q_T \cdot b + Ef + \left[ (E'-E) + \frac{1}{4} f'^2 E
-\frac{1}{2}\, q_T f' \right] \widetilde{U} \, , \label{phaseAS}
\eeq
where $\widetilde{U} = 2p^u\nu +U_0$.

The result (\ref{phaseAS}) can  also be written as
\beq
- q_T \cdot b + Ef + \frac{\widetilde{U}}{4E}
\Bigl[ q_T - f' E \Bigr] ^2
\eeq
by means of the mass-shell condition
\beq
q_T^2 = 4E(E'-E) \, .
\eeq

The above calculation can be repeated, of course, for any initial
transverse vector $x_T(\nu_i) = (b_1,~b_2)$.
The Green function is therefore
\bea
\lefteqn{G_c (p,p'/AS) =\mbox{Lim} \frac{2p^u \delta(q_u)}{\nu} \int_{0}^{\nu}
\di \nu' \int \di^2 b \: \exp \Bigl[ -iq_T \cdot b + i E f(b) \Bigr]} \nnu
\\
& & \times \exp \left[i\nu' \left( q_T - f'E \right)^2 \right] \, ,
\eea
which is dominated by a stationary phase at
\beq
q_T = Ef' \, , \label{st_ph_AS}
\eeq
corresponding to the known  \cite{FPV} relation between scattering angle
and impact
parameter in the generalized AS metric:
\beq
\tan  \vartheta /2 = - f'/2 \, .
\eeq
 In that saddle approximation we can also use:
\beq
\lim_{\nu \rightarrow \infty}\frac{1}{\nu} \int_0^\nu \di \nu' \exp \left[
 i\nu' \left( q_T - f'E \right)^2 \right] =
1 \, ,
\eeq
and therefore obtain for the scattering amplitude
\beq
G_c (p,p'/AS) =4E \delta(q_u)
\int \di^2 b \: \exp \Bigl[ - iq_T \cdot b + i E f(b) \Bigr] \, ,
\eeq
in agreement with the known result~\cite{tHooft}.

\subsection{Scattering by a black hole}

\quad The problem of the scattering of a massless scalar particle
by a  (Schwarzschild) black hole  is most easily discussed
in  spherical  coordinates.
We introduce the Green function for the partial waves $(E,l,m)$ and
$(E',l',m')$ as
\beq
G(E,l,m,E',l',m') = \int \di^4 x \int \di^4 y \langle l'm'E'|y \rangle
\langle x|lmE \rangle G(x,y)
\eeq
for which, in standard notation,
\bea
\langle x|lmE \rangle  &=& Y_{lm}\left( \phi,\, \theta \right)
 R_{El}\left( r \right) \exp -iEt \nnu \\
\langle l'm'E'|y \rangle  &=& Y_{l'm'}^*\left( \phi,\, \theta \right)
 R_{E'l'}^*\left( r \right) \exp -iE't \, ,
\eea
in order to transform to spherical  coordinates.

 In this case the propagator for the partial waves can be written as
\bea
\lefteqn{G(E,l,m,E',l',m') =  \lim_{\nu_f \rightarrow +\infty \atop
\nu_i \rightarrow -\infty} \frac{1}{\nu_f - \nu_i}
 \int \di X^\mu(\nu_i) } \nnu \\
&& \times \exp \left[ -iEt(\nu_i) + iE't(\nu_f) \right]  \nnu \\
& &
\times
Y_{lm}\left( \phi (\nu_i), \theta (\nu_i) \right) Y_{l'm'}^* \left( \phi
(\nu_f) , \theta (\nu_f) \right)
 R_{El}\left( r(\nu_i) \right) R_{E'l'}^* \left( r(\nu_f) \right) \, ,
\eea
where the measure is
\beq
\di X^\mu = \di t(\nu_i)\, r^2(\nu_i)\, \di r(\nu_i) \,\di\phi(\nu_i)
\di \cos\theta(\nu_i)
\eeq
and one
integrates over the initial conditions, the final ones being given by the
dynamics.

The classical equations of motion are given in parametric form
as~\cite{Darwin}:
\bea
\left( \frac{\di r}{\di \nu} \right)^2
&=& 1 - \frac{b^2}{r^2} \left( 1 - \frac{r_0}{r} \right)
\, , \nnu \\
\frac{\di\theta}{\di \nu} &=& \frac{b}{r^2} \, , \nnu \\
\frac{\di t}{\di \nu} & =& \left( 1 - \frac{r_0}{r}
\right)^{-1} \, , \label{motion}
\eea
where $b=l/E$ and $r_0 = 2MG$ is the horizon radius.

We consider a planar motion with $m=0$ and
\beq
\phi(\nu_i) = \phi(\nu_f) \, ,
\eeq
which implies  that the integration over the azimuthal initial angle
gives a
$\delta_{0,m'}$.

The integration over the angle $\theta(\nu_i)$ can also be performed
directly, to yield
\beq
\delta_{ll'}\, i^{2l} P_l \left( \cos \Delta \theta \right)\, ,
\eeq
where
\beq
\Delta \theta = \theta(\nu_i)  - \theta(\nu_f) \, . \label{theta}
\eeq
 Similarly, we can integrate over the initial times to obtain
 \beq
 \int \di t(\nu_i) \exp \Bigl[ -i (E - E') t(\nu_i) +
 iE' \Delta t \Bigr] =
 \delta(E-E') \exp \Bigl[ iE \Delta t \Bigr] \, ,
 \eeq
 where
\beq
\Delta t = t(\nu_f) - t(\nu_i) \, . \label{t}
\eeq
 In this way we find:
 \bea
 \lefteqn{G(l,E) = \lim_{\nu_f \rightarrow +\infty \atop
\nu_i \rightarrow -\infty} \frac{1}{\nu_f - \nu_i} \int  r^2(\nu_i)
\di r(\nu_i)
 i^{2l}\exp \Bigl[ iE\Delta t \Bigr]} \nnu \\
 & &  \times
  P_l \left( \cos \Delta \theta \right)
   R_{El}\left( r(\nu_i) \right) R_{El}^* \left( r(\nu_f) \right) \, ,
 \eea
 which, for large $l$, can be approximated by
 \bea
\lefteqn{G(l,E) = \lim_{\nu_f \rightarrow +\infty \atop
\nu_i \rightarrow -\infty} \frac{i^{2l}}{\nu_f - \nu_i} \int
\frac{r(\nu_i) \di r(\nu_i)}{r(\nu_f)} }  \nnu \\
& & \times \exp i \Bigl[ E  \Delta t
 - p_r r(\nu_f) - p_r' r(\nu_i) -l\pi \Bigr]
 \nnu \\
 & &   \times \left[ \exp \left( il\Delta \theta + i\frac{\pi}{4} \right)
- \exp  \left( -il\Delta \theta - i\frac{\pi}{4} \right) \right] \, ,
\label{gg}
\eea
where we have taken, respectively, an incoming wave  at the
initial  time and an outgoing wave  at the final time. In our
parametrization, and for $\nu_i$ and $\nu_f$ sufficiently large, we
have   $r(\nu_i) \simeq r(\nu_f) =L$, in such a way that the
argument of the exponential  no longer depends
 on $r(\nu_i)$. Therefore, the
remaining integration over $r(\nu_i)$ can be performed to give
\beq
 \frac{1}{\nu_f-\nu_i} \int \di r(\nu_i) \simeq
\frac{1}{\nu_f-\nu_i} \int_{\nu_i}^{\nu_f} \di L \longrightarrow
1 \, ,
\eeq
 and $2L \simeq \nu_f-\nu_i$.
 The radial momenta are
\beq
p_r \simeq p_r' = E\sqrt{L^2 -b^2}/L \simeq E L \, .
\eeq
The partial wave amplitude (propagator) $S(l,E)$ is thus given by:
\beq
S(l,E) = G(l,E) = \exp (2i \delta_l)
\eeq
with:
\beq
2\delta_l = \frac{E}{\hbar} \left[\Delta t -2 L \right]
  \pm
  \left[ l \left(  \Delta \theta +\pi \right) + \frac{\pi}{4} \right]
   \label{delta}
 \, .
\eeq

Equation (\ref{delta}) is the main result of this section. It
expresses the phase shifts in terms of $\Delta t$ and $\Delta \theta$,
 which are defined
in (\ref{t}) and (\ref{theta}) and can be obtained
by integrating the classical equations of motion
(\ref{motion}).

 The free-motion phase shift is just
\beq
2\delta_l^{(0)} = \pi l +\frac{\pi}{4} \, .
\eeq

Going back to momentum space,
the propagator can now be written as
\beq
G(E,\, \vartheta /Schw) = \sum_l \left( 2l+1 \right) P_l (\cos \vartheta )
 G(l,E)
\eeq
where $\vartheta$ is the scattering angle.

The sum over $l$ is dominated by a stationary phase at
\beq
\pm \vartheta = \frac{\di \delta_l}{\di l} \label{st-ph} \, .
\eeq

 By using the explicit form of $\delta_l$, (\ref{delta}), we find
the stationary
 phase condition (\ref{st-ph}) in the form ($L \rightarrow \infty$):
 \beq
 \pm \vartheta = b \int_{r_{min}}^\infty \frac{\di r}{r^2\sqrt{1 - \left(1
 - r_0/r \right) (b^2/r^2)}}  - \pi \, ,
\eeq
 which
agrees with the classical equation for the orbit found
in~\cite{Darwin} in terms of an elliptic integral of the first kind.

The shift in the
time coordinate in (\ref{delta}) is particularly interesting. It
 can be written as
\beq
\Delta t = \nu_f - \nu_i + 2r_0 \int_{r_{min}} \frac{\di r}{r - r_0}
\frac{1}{\sqrt{1- (1-r_0/r) (b^2/r^2)}} \, .
\eeq
We see that for $r_{min} \le r_0$ (corresponding
 to angular
momenta $l < E\tilde{b}$, where $\tilde{b}=3\sqrt{3}r_0/2$ is the
critical impact parameter below
which classical capture takes place), $\Delta t$ acquires an imaginary
part:
\beq
\mbox{Im} \,\Delta t = 2\pi r_0 \, . \label{im}
\eeq
Thus the incoming wave is almost completely absorbed for $b
 <
\tilde{b}$ (recall that, in order for our approximations to be valid,
we are always at $r_0 E >> 1$), the
transmission probability being  ${\cal O} \left( e^{-4\pi r_0 E} \right) =
e^{-E/T_H}$, where
\beq
T_H = 1 /4\pi r_0 \, ,
\eeq
is the black hole temperature.

 Finally, the imaginary part (\ref{im}) gives an inelastic  contribution to
   partial wave cross sections with $l < E\tilde{b}$ :
\beq
 \sigma_{in}^l =  \frac{\pi}{E^2} (2l+1) \biggl[
1 - \exp\left( -E/T_H  \right) \biggr]  \, .
\eeq

 The above results are in full agreement
with those obtained in the literature by a more direct (and
involved) computation~\cite{Sanchez}.
In particular, the total inelastic cross section can easily be
evaluated to give
\beq
\sigma_{in} = \frac{\pi}{E^2} \sum_{l=0}^{l=E\tilde{b}} (2l+1)
\simeq \frac{27}{4} \pi r_0^2 \, .
\eeq

In closing this section we wish to mention that sometime ago  Hartle
and  Hawking~\cite{H-H} addressed the problem of spontaneous black hole
radiation by using an approach where complex trajectories dominate the
path integral.

\section{Two-Body Collisions}
\setcounter{equation}{0}

\qquad After having checked our method in the case of scattering in an
external field, we shall now turn to our real goal, the Planckian-energy
collision of two light (massless) quanta.

As explained elsewhere~\cite{(1),(6)}, the semiclassical approximation should
be valid in the Planckian energy regime for any finite value of the parameter
$R/b$ and provides an ${\cal S}$-matrix of the form:
\beq
 \ln {\cal S}(E,b) = \frac{i G s}{\hbar} F(R/b,\log s,
\log b) \Bigl( 1 + {\cal O} (\hbar) \Bigr) \, . \label{555}
\eeq

This approximation corresponds to resumming
 all Feynman graphs which become (possibly disconnected) tree diagrams once the
 scalar-particle propagators are cut. They are illustrated in figs. 1 and
 2 for the case   of elastic and inelastic scattering, respectively.
Notice that
neither matter loops nor pure  graviton loops are allowed
in these diagrams. Simple
power counting arguments show~\cite{(1),(6)} that these loops  only
contribute to the ${\cal O} (\hbar)$ corrections indicated in~(\ref{555}).

In the functional integral, the semiclassical limit corresponds
 to a saddle point approximation in which the classical solution
for $X^{\mu}$ and $g_{\mu\nu}$ are inserted in the action. The
 basic idea is thus quite
simple: use the homogeneity in $g_{\mu\nu}$ of the bulk terms in
 the gravitational and scalar particle
 actions   to reduce the whole action on the classical equations of
motion to just surface terms (the gravitational term $S_K$ and the
term $S_{p_{ext}}$).

 We first discuss the case of elastic scattering,
reproducing in a simple and elegant way
the known leading-eikonal result, and then briefly discuss
how the method retains its validity beyond such approximation.

 \subsection{Elastic Scattering}

\qquad In the absence of matter loops,
the elastic scattering amplitude is given by
\bea
\lefteqn{ (2\pi)^4
\delta^{(4)} \left( p_1 +p_2 - p_1' -p_2' \right)
 A \left(p_1 + p_2 \rightarrow p_1' + p_2' \right)
   =  } \nnu \\
& & \times \int [\di g_{\mu\nu}] \exp \Bigl( iS_{g} \Bigr)
 G_c^1(p_1,p_1'/g)G_c^2(p_2,p_2'/g) \, . \label{A0}
\eea
where, for the sake of simplicity, we restrict our
attention to the case of two distinguishable scalar particles,
 thus avoiding complications with crossed or annihilation diagrams.

By means of (\ref{Gc}),
the right-hand side of eq.~(\ref{A0}) can be written as
\beq
\lim_{\nu_i \rightarrow \infty \atop
p_i^2 \rightarrow 0} \left(\frac{-i}{\nu_1} \right)
\left(\frac{-i}{\nu_2} \right)
\int [\sqrt{-g(X_1)}
\di X_1^\mu] [\sqrt{-g(X_2)} \di X_2^\mu][\di g_{\mu\nu}] \exp \Bigl(i S
\Bigr)
\label{A1}
\eeq
for
\beq
S= S_g + S_1 + S_2 + S_{p_{ext}} \, , \label{S-tutti}
\eeq
where
\bea
S_g &=& -\frac{1}{2\kappa^2} \int \sqrt{-g}\: {\cal R}\: \di ^4 x +
S_K \, , \nnu \\
 S_i &=& \frac{1}{2} \int_{0}^{\nu_i} \di \tau_i\:
g_{\mu\nu}  (X_i)\dot{X}_i^\mu (\tau_i)\dot{X}_i^\nu (\tau_i) \, , \nnu \\
S_{p_{ext}} &=& p_1 \cdot X_1(0) -  p_1' \cdot X_1(\nu_1) +
 p_2 \cdot X_2(0) -  p_2' \cdot X_2(\nu_2) \, ,
 \eea
 and $S_K$ is the surface term (\ref{2.16}).

 The classical equations of motion are obtained by a
variation with respect to $g_{\mu\nu}$:
 \beq
 {\cal R}^{\mu\nu}(x) - \frac{1}{2}\, g^{\mu\nu} {\cal R}(x)  =
-\frac{\kappa^2}{2\sqrt{-g}}
 \sum_{i=1,2} \int_{0}^{\nu_i} \di \tau_i \:\delta^{(4)} \left( x -
X_i(\tau_i)
\right)  \dot{X}_i^\mu (\tau_i)\dot{X}_i^\nu (\tau_i)
\eeq
and with respect to $X^\mu_i$:
\beq
\ddot{X}_i^\mu + \Gamma^\mu_{\rho\sigma} (X_i) \:\dot{X}_i^\rho
\dot{X}_i^\sigma =0 \qquad \mbox{for} \quad i=1,2.
\eeq

As we have already discussed in the case of the propagation in an
external field, the homogeneity of $S_i$ and, now, of $S_g$ as well,
yields
an action that reduces, on the equations of motion, to a surface term:
 \beq
 S \Bigl| _{sol'n} = S_K + S_{p_{ext}}  \Bigr. \, . \label{surf}
 \eeq

In the semiclassical limit, we therefore have
\bea
\lefteqn{ (2\pi)^4 \delta^{(4)} \left( p_1 +p_2 - p_1' -p_2' \right)
 A \left(p_1 + p_2 \rightarrow p_1' + p_2' \right)
   = }\nnu  \\
& & \lim_{\nu_i \rightarrow \infty \atop
p_i^2 \rightarrow 0} \frac{-1}{\nu_1\nu_2}
\int \di X_1(0) \di X_1(\nu_1) \di X_2(0) \di X_2(\nu_2) \exp \left( i
S|_{sol'n} \right) \, . \label{A2}
\eea

The amplitude (\ref{A2}) can be rewritten by introducing the variables
\bea
y_i &=& X_i(0) - X_i(\nu_i) \nnu \\
\Delta & = &  \left( X_1(0) +X_1(\nu_1) \right)/2 -
\left(  X_2(0) + X_2(\nu_2) \right)/2  \nnu \\
q &=& (p_1-p_1')/2 - (p_2-p_2')/2 \nnu \\
P_i &=& (p_i+p_i')/2\nnu \\
\Sigma &=& \sum_i X_i
\eea
after which
an integration  over $\Sigma$ cancels the overall
$\delta$-function on the left-hand side of eq.~(\ref{A2}).
We thus obtain
\bea
\lefteqn{ A \left(p_1 + p_2 \rightarrow p_1' + p_2' \right) = } \nnu \\
& & \lim_{\nu_i \rightarrow \infty \atop
p_i^2 \rightarrow 0} \frac{-1}{\nu_1\nu_2}
\int \di^4 \Delta\: \exp  \Bigl[ i q \cdot \Delta + iP_1 \cdot y_1
+  iP_2 \cdot y_2  + iS_K \Bigr] \, .\label{ampl}
\eea

It is at this stage that (possibly non-perturbative) approximate solutions
for the classical collision geometries or their generalizations should be
used to calculate the amplitude (\ref{ampl}). In this paper we only carry out
a first non-trivial check of the method by showing how the known leading
eikonal result arises from our surface terms. The derivation is then
compared with the one using perturbation theory in the Newton constant.

\subsection{Leading Eikonal Approximation}

\qquad The leading eikonal approximation corresponds, in our approach, to
working out the surface terms appearing in (\ref{ampl}) to ${\cal O} (E^2)$.
For the external particle contribution the calculation is simple since, to
this order, one just needs the shifts $y_i$ to ${\cal O} (E)$. To this
purpose we can use the leading order form of the metric which consists of the
two independent AS shock waves created by the two particles:
\bea
\lefteqn{h_{\mu\nu} = h_{\mu\nu}(1) +  h_{\mu\nu}(2) }  \nnu \\
& & = - \frac{\kappa^2 E_1}{2\pi}\, \tilde{\nabla}_\mu U \,
\tilde{\nabla}_\nu
U \,
 \delta (U) \ln \left( y^2 + z^2 \right) \nnu \\
& & \quad - \frac{\kappa^2 E_2}{2\pi}  \, \tilde{\nabla}_\mu V \,
\tilde{\nabla}_\nu
V \,
 \delta (V) \ln \left( (y - b)^2 + z^2 \right) \, ,
\label{case}
\eea
where, in suitable coordinates, the transverse vector $b$ can be
 identified  with
the transverse components of the four-vector $\Delta$ appearing in
(\ref{ampl}).

The calculation of the shifts is now identical to the one
used in section 4.2, in the case of scattering from an external field.
In a generic boosted frame, in which the two particles
are coming against each other along the $x$ axis, the initial momenta are
\beq
 p_1' \simeq p_1 = (E_1,E_1,0,0)
 \eeq
 and
 \beq
 p_2' \simeq p_2 = (E_2,-E_2,0,0)
 \eeq
 and we readily find (see eq.(\ref{shock}):
 \beq
P_1 \cdot y_1 =P_2 \cdot y_2= - \frac{\kappa^2 E_1 E_2}{2 \pi} \,
 \ln b^2 \, .
\eeq

We see that the contribution of the external momenta is twice the desired
value because, unlike in the external metric problem, there are now two
  identical contributions from each colliding particle.
 If the known result is to emerge, the gravitational surface term
$S_K$ should contribute as:
\beq
S_K  = \frac{\kappa^2 E_1~E_2}{2 \pi} \, \ln b^2 \, , \label{corr}
\eeq

 We shall first show how this result arises very simply from the Bondi
masses and next compare it with the perturbative calculation.
An apparent difficulty in both approaches is that we need the action
to ${\cal O} (E^2)$ whereas the metric (\ref{case}) is valid only to
to ${\cal O} (E)$. Nevertheless, as we shall see, having the metric
 to ${\cal O} (E)$ is actually sufficient.

We recall, from the appendix, the meaning of the Bondi masses.
 $M_{-}(v)$,  at $\cal{I}^-$, is the energy brought
into the system by  the (advanced) time $v$, while
 $M_{+}(u)$,  at  $\cal{I}^+$, is the energy present in the system at
(retarded) time $u$. Note the difference in notation and meaning:
the arguments $v$ and $u$ of the Bondi masses are exactly null
variables and label,  respectively, converging
and diverging spherical congruences, whereas
 the $V$ and $U$ coordinates appearing in (\ref{case}) are
only approximately null (they are such only in the flat metric) and label
plane congruences.

To lowest order in $E_1$, $E_2$, the only relevant effect is again
the above mentioned shift of the geodesics computed this time in terms
of the variables $u$, $v$. Consider first the case of a single
non-interacting massless particle. In this case, $U$ remains
exactly null and the Bondi masses are of the form:
\beq
M_{-}(v) =  E_1  ~\theta (v) ~ \, , \, ~
M_{+}(u) =  E_1  ~\theta (-u) \, , \, ~  (E_2 = 0) \, , \label{free}
\eeq
expressing the fact that a particle of energy
 $E_1$ comes in through $\cal{I}^{-}$
at the instant $v=0$ and goes out through $\cal{I}^{+}$ at the instant $u=0$.
There are no further energy fluxes either through $\cal{I}^{-}$ or through
$\cal{I}^{+}$.

In the case of two particles, there are still
no interactions at past null infinity. Therefore the Bondi mass $M_{-}$ is
simply
\beq
M_{-}(v) =  E_1   ~\theta (v) + E_2   ~\theta (v) \, ,  \label{mminus}
\eeq
which means that both particles show up at $\cal{I}^{-}$ at the instant $v=0$.

The time instants of their arrival at $\cal{I}^{+}$ will now be shifted
 in comparison with (\ref{free}).
By using the geodesic shifts in $U$ and $V$ and the
     asymptotic
 relation  between the two sets of coordinates, it is easy to show
that the first particle will arrive at $\cal{I}^{+}$ at $u=u_1$ and the
second at $u=u_2$ where
\bea
u_1 &=& -\frac{\kappa^2 E_2}{2\pi} \ln b^2 \, ,\nnu \\
 u_2 & =&  -\frac{\kappa^2 E_1}{2\pi} \ln b^2 \, . \label{ushifts}
\eea

The Bondi mass at $\cal{I}^{+}$ is, therefore,
\beq
M_{+}(u) =   E_1~\theta (u_1 - u)   + E_2~\theta (u_2 - u)  \, .  \label{mplus}
\eeq

 It should be stressed here that, while (\ref{mminus}) is an exact expression,
(\ref{mplus}) accounts only for the lowest-order effect of
the time shift and neglects higher-order contributions from gravitational
 radiation and
mass renormalization.

Inserting eqs.(\ref{mminus}) and (\ref{mplus}) into the expression (\ref{2.16})
for $S_K$:
\beq
S_K = - \frac{1}{2} \int_{-\infty}^{+\infty} \di x \Bigl[ M_+(-x) - M_-(x)
\Bigr]
\, ,\eeq
we obtain precisely the desired result (\ref{corr}). Note that
the time shifts (\ref{ushifts}) are to be computed from the metric only
 to ${\cal O} (E)$ since there is already a factor $E$ in the Bondi masses.
This is the way a metric of ${\cal O} (E)$ leads
 to an action of ${\cal O} (E^2)$. With Bondi masses limited to first order in
$E$,
one would have for $M_+$ an expression similar to (\ref{mminus})  and a
vanishing result for $S_K$, in complete agreement with the fact that there is
no
contribution linear in $E$ in the action.

The  perturbative calculation proceeds instead as follows.
  The surface
term $S_K$ is transformed back to the volume integral of a total derivative
(see (\ref{2.13}) after using Gauss' theorem) yielding
\beq
S_K = \frac{1}{2\kappa^2} \int_{\Omega}
\di ^4 x \sqrt{-\tilde{g}} \left( \tilde{\nabla}^2 \tilde{g}^{\mu\nu}
h_{\mu\nu} -\tilde{\nabla}^\mu\tilde{\nabla}^\nu h_{\mu\nu}
\right)\, , \label{5.12}
\eeq
where the tilde stands for flat space-time.

 For a direct calculation
of (\ref{5.12}) we would need a solution to second order in $E$ since, again,
the insertion  of the lowest-order solution (\ref{case}) in (\ref{5.12})
yields zero.  This problem  can  be avoided,
however,
by  noticing that for a traceless source---such as our two massless particles--
\beq
 {\cal R} =  \left(\tilde{\nabla}^2 \tilde{g}^{\mu\nu}
-\tilde{\nabla}^\mu\tilde{\nabla}^\nu \right) h_{\mu\nu} +
{\cal O}(h_{\mu\nu}^2) = 0
\eeq
and therefore $S_K$ can be written as the quadratic part of the
Hilbert action plus higher-order terms, that is
 \bea
\lefteqn{S_K  =  \frac{1}{8\kappa^2} \int_{\Omega}
\di ^4 x \sqrt{-\tilde{g}} h_{\mu\nu} \left[
\tilde{g}^{\mu\alpha}\tilde{g}^{\nu \beta}\tilde{\nabla}^2
+ \tilde{g}^{\alpha\beta}\tilde{\nabla}^\mu\tilde{\nabla}^\nu
\right. }\nnu
\\
& & \quad \left.
-2\tilde{g}^{\mu\alpha}\tilde{\nabla}^\nu\tilde{\nabla}^\beta -
\tilde{g}^{\mu\nu}\tilde{g}^{\alpha \beta}\tilde{\nabla}^2
+ \tilde{g}^{\mu\nu}\tilde{\nabla}^\alpha\tilde{\nabla}^\beta
\right] h_{\alpha\beta} + {\cal O}(h_{\mu\nu}^3) \, . \label{SK}
\eea

The metric (\ref{case}) is now sufficient because (\ref{SK}) is quadratic
in the solution. We thus obtain that
\beq
S_K  =    \frac{1}{8\kappa^2} \int \di^4 x \sqrt{-\tilde{g}} \:
h_{\mu\nu}
\left( \tilde{g}^{\mu\alpha}\tilde{g}^{\nu \beta}\tilde{\nabla}^2
\right)  h_{\alpha\beta}
 =  \frac{\kappa^2 E_1~E_2}{2 \pi} \, \ln b^2 \, , \label{op}
 \eeq
since in this case
\beq
\tilde{g}^{\alpha\beta}h_{\alpha\beta} = \tilde{\nabla}^\alpha
h_{\alpha\beta} =0 \, .
\eeq

We have thus obtained, by two conceptually different calculations, the same
final result for the surface action $S_K$. Comparison of the two methods
 confirms the validity of our approach. It also shows, in our opinion,
that working in terms of the Bondi masses is simpler and physically more
meaningful. The
crucial test will come, of course, with
non-perturbative computations, something we do not want to attempt here.

The scattering amplitude (\ref{ampl}) can now be computed.
   In order for
the two particles to scatter at all, we have to impose the two conditions
\beq
X^-_2 (\nu_2) > X^-_1(0) \, , \qquad X^+_1(\nu_1) > X^+_2 (0) \, .
\eeq
This gives two $\theta$-functions that  can be written in terms of the
variable $\Delta$ as
\beq
\theta \left( \Delta^+ -E_1 \nu_1 \right) \theta \left( \Delta^- -E_2 \nu_2
\right) \, .\label{cond}
\eeq
The condition (\ref{cond}) enforces the requirement that between $\nu = 0$ and
$\nu = \nu_1$, as well as between $0$ and $\nu_2$, the two trajectories
cross each other.
 The integration over the light-like
components of $\Delta$ thus gives \bea \lefteqn{ \int \di \Delta^+ \di
\Delta^- \exp \left[ {i\over 2} q_+ \Delta^+ +{i\over 2} q_- \Delta^- \right]
 \theta \left(
\Delta^+ -E_1 \nu_1 \right) \theta \left( \Delta^- -E_2 \nu_2  \right) = }\nnu
\\ & & 4 \left(\frac{\exp  \left({i\over 2}q_+ E_1 \nu_1 \right) -
1}{iq_+}\right)\,\,
\left(\frac{\exp  \left({i\over 2}q_- E_2 \nu_2 \right) - 1}{iq_-}\right)\,\,
\:
{\buildrel { q_+,q_- \rightarrow 0} \over \longrightarrow}\:
 E_1~E_2 \nu_1\nu_2
\eea
canceling the factor $1/\nu_1\nu_2$ in (\ref{A2}). We
therefore obtain:
 \beq
A \left(p_1 + p_2 \rightarrow p_1' + p_2' \right) =
E_1 E_2
\int \di^2 b \exp  \left[ i q_T \cdot b
 -i \frac{\kappa^2 E_1 E_2}{2 \pi} \, \ln b^2 \right] \, ,
\label{Amp}
\eeq
in agreement with previous results~\cite{(1),(3),(4),(5),Bellini}.

\subsection{Inclusion of Gravitational Bremsstrahlung}

\qquad In order to extend the calculations of the previous section
to
higher orders in $R/b$, one has to deal with the problem
of gravitational bremsstrahlung (see fig. 2). Equivalently, in
the language of  particle physics,
one encounters
  infrared divergences in amplitudes with a given number of final
gravitons and some Block-Nordsiek-type  treatment becomes necessary
in order to be able to extract finite, physically meaningful
results~\cite{(6)}.

The general reasoning given in section 2.2 insures that
the surface nature
of the semiclassical $\cal S$-matrix remains valid even after allowing for the
effects of gravitational radiation. It is however instructive to see how this
works in detail in the case of two-body inelastic collisions  within a
conventional LSZ treatment of the $\cal S$-matrix.

According to LSZ the  amplitude for two scalar particles ($\psi$) to
scatter with emission of $n$ gravitons ($h$) is given (symbolically) by
\beq
A(2 \rightarrow 2 + n) = \Bigl. \langle \psi \, \psi \, \psi
\, \psi \, h\, h\, \cdots \,h \rangle \Bigr| _{\mbox{\small LSZ}} \label{LSZ}\,
, \eeq
that is by applying the standard LSZ steps (Fourier transform, truncation
of propagators and going on shell) to the $4+n$-point function shown in
 (\ref{LSZ}).

These amplitudes are generated by the functional
\beq
  A(2 \rightarrow 2 \, ,\: J) = \Bigl. \langle \psi \, \psi \, \psi
\, \psi \, \exp (h J) \rangle \Bigr| _{\mbox{\small LSZ}} \, , \label{LSZ2}
\eeq
through standard functional differentiation with respect to the
external source $J$.
In the absence of scalar-field loops, the LSZ operation amounts to
replacing pairs of scalar fields by the corresponding amputated propagators
$G_c$ and  the source $J$ by $\Delta^{(2)} h_{as}(x)$,
where $\Delta^{(2)}$ is the quadratic operator appearing in
eq.~(\ref{op}). $h_{as}(x)$ then becomes the new external source.

 Finally, in order for the produced gravitons to be on-shell, the condition
\beq
\Delta^{(2)} h_{as}(x)=0   \label{(5.24)}
\eeq
has to be imposed at the end of the computation. This
allows us to replace $h(x)$ in eq.~(\ref{LSZ2}) with the full
$g_{\mu\nu}(x)$.

 We conclude that  a power expansion in $h_{as}$ of the
${\cal S}$-matrix functional
\beq
{\cal W} (h_{as}) = \int [\di g_{\mu\nu}]
 \exp \left[ iS_{g} +iS_{br} \right]
 G_c^1(p_1,p_1'/g)G_c^2(p_2,p_2'/g) \, ,  \label{(5.23)}
\eeq
where
\beq
S_{br} = \int \di ^{4} x \: g(x) \,\Delta^{(2)} h_{as}(x) \, ,
\label{(5.22)}
\eeq
    gives the
 amplitudes for emission
of any number of gravitons. Following the procedure of Section 5.1, we can
rewrite (\ref{(5.23)}) as a functional integral over $g_{\mu\nu}$, $X_1^\mu$
and
$X^\nu_2$ with a modified action $S$ containing, beside the terms in
(\ref{S-tutti}), the additional bremsstrahlung term (\ref{(5.22)}).

The crucial  observation at this point is that
the additional term ({\ref{(5.22)})  is still effectively
homogeneous of degree one in $g_{\mu\nu}$. Therefore, within the stated
approximations,
we are still left with the result
(\ref{surf}) and
\beq
{\cal W}(h_{as}) = \exp \left[ i \left. (S_K + S_{p_{ext}}) \right| _{sol'n}
\right] \, , \label{W-as}
\eeq
as expected from the general arguments of section (2.2)

Note that, while the two surface terms occurring in (\ref{W-as}) are still
the same ones as in the
elastic case, their actual values now
 include,  implicitly, a dependence on gravitational radiation through
$h_{as}(x)$.
Giving $h_{as}(x)$ amounts to specifying a definite final coherent state of
gravitons, the closest one can get quantum-mechanically  to a classical
radiation field. For a  generic
  $h_{as}(x)$, one will still be plagued by infrared problems.
However, as in the analogue QED problem, a well-defined coherent state
 for which infrared divergences cancel will
be chosen by the dynamics and will represent  quantum mechanically
the classical asymptotic gravitational radiation field generated in the
collision.

In concluding this section, we want again to remark that the validity of
(\ref{surf}) does not depend on any approximation except the tree
approximation, for both the gravitational field and the first-quantized
particles. The surface term has one and the
same form for any interactions of the gravitating fields but its value is
different
for different interactions.

To what extent the surface nature of gravity persist when graviton loops are
taken into account is discussed in the next section.

\section{Inclusion of graviton loops}
\setcounter{equation}{0}

\qquad The extension of the above ideas
beyond the classical approximation for the metric is based on
\cite{canada,aaa,GHP} which dealt with the quantum fluctuations of the
conformal mode of the gravitational field. In this section
we first give a new, more direct,
derivation of the result of these works that also shows at which points
care
should be taken when using these results; next, we discuss its relevance
for the present work.

 Let  us write
\beq
g_{\mu \nu} (x) \,=\, \bar{g}_{\mu \nu}(x) \phi^{2}(x) \, , \label{6.1}
\eeq
 singling out the conformal mode of
the
gravitational field,
and impose on $\bar{g}_{\mu \nu}(x)$ a constraint
\beq
\chi(\bar{g}(x))\,=\,0  \, ,\label{6.2}
\eeq
such that the equation
\beq
\chi(\Omega^{2}(x) \bar{g}(x))\,=\,0 \, , \label{6.3}
\eeq
with $\bar{g}$ satisfying (\ref{6.2}), has only one solution
\beq
\Omega^{2} (x)\,=\, 1. \label{6.4}
\eeq
We may then reparametrize the ten components of $g_{\mu \nu}$ into nine
independent components of $\bar{g}_{\mu \nu}$, constrained by condition
(\ref{6.2}), and the conformal mode $\phi(x)$.

With respect to $\bar{g}_{\mu \nu}$, eq. (\ref{6.2}) plays the role of a
gauge-fixing condition, which removes the arbitrariness of local conformal
transformations $\bar{g}_{\mu \nu} \rightarrow \Omega^{2} \bar{g}_{\mu \nu}$.
We shall consider only the transformations that become the identity at the
asymptotically flat infinity; the boundary conditions for the conformal mode
are
\beq
\phi (x) \left|_{~\atop {{\cal I}^{-} , {\cal I}^{+}, I^{\circ}}}\,=\,1
\right. \, . \label{6.5}
\eeq
The choice of the gauge-fixing function $\chi (\bar{g})$ is limited only
by the requirement that the continuous matrix
\beq
\bar{Q}(x,y)\,=\, \bar{g}_{\mu \nu}(x) \,\frac{\delta \chi (\bar{g}(y))}
{\delta \bar{g}_{\mu \nu}(x)}  \label{6.6}
\eeq
be invertible.

Our goal is to rewrite the functional integral (\ref{2.19}) in terms of a set
of variables containing the conformal mode $\phi (x)$ explicitly.
However, instead of introducing $\phi$ and the nine independent components of
$\bar{g}_{\mu \nu}$, we introduce  eleven variables---$\phi$ and the ten
components of $
\bar{g}_{\mu \nu}$---and insert the delta-function $\delta(\chi(\bar{g}))$ in
the measure. By calculating the relevant Jacobians, we find
\beq
\prod_{x,\mu \leq \nu} \di g_{\mu \nu}\,=\,(\det \bar{Q}) \prod_{x}\, \phi^{19}
\delta(\chi ( \bar{g})) \di \phi \prod_{\mu \leq \nu} \di \bar{g}_{\mu \nu} \,
,  \label{6.7}
\eeq
with $\bar{Q}$ given by (\ref{6.6}). The measure in (\ref{2.19}) contains local
factors in  $g_{\mu \nu}$ \cite{FV} which, after the substitution of
(\ref{6.1}), will change the total power of $\phi$ in the transformed measure.
However, in what follows, we shall ignore the local factors in the measure
altogether by assuming, for instance, dimensional regularization.

Next, let us find the form of the action (\ref{2.3}) in the new variables. We
have
\beq
\frac{1}{6}\,
\sqrt{-g}\,{\cal R}(g)\,=\, \sqrt{-\bar{g}} \left( \phi \bbox \phi
\,+\,\frac{1}{6}\,{\cal R} (\bar{g})\phi^{2} \right)  \, , \label{6.8}
\eeq
where the quantities and operators with a bar refer to $\bar{g}_{\mu \nu}$.
The transformation of the boundary term $S_K[g]$ is also non-trivial, and
its effect is that the total action contains no second derivatives of $\phi$:
\bea
 \lefteqn{- \frac{1}{2 \kappa ^2} \int \di^{4}x \sqrt{-g}\,
{\cal R}\,+\,S_K[g] =} \nnu \\
& & - \frac{3}{\kappa ^2}
\int \di^{4}x \sqrt{-\bar{g}} \left[-(\bar{\nabla}
\phi)^{2}\,+\, \frac{1}{6}\,{\cal R} (\bar{g}) \phi^{2}
\right]\,+\,S_K[\bar{g}]
\label{6.9}
\eea
so that its variational equation in $\phi$ is of the form
\beq
\left( \bbox\,+\, \frac{1}{6}\, {\cal R} (\bar{g}) \right) \phi\,=\,0 \, .
\label{6.10} \eeq

As a result, the functional integral (\ref{2.19}) for the case of pure gravity
takes the form
\bea
\lefteqn{{\cal S} =  \int [\di \bar{g}_{\mu \nu}] [\di \phi ] (\det \bar{Q})\,
\delta  (\chi(\bar{g}))\, \exp \left\{- \frac{3i}{\kappa ^2} \int \di^{4} x
\sqrt{- \bar{g}}  \left[
- (\bar{\nabla} \phi)^{2} \right. \right. }\nnu \\
& & \left.\left. +\: \frac{1}{6} \, {\cal R} (\bar{g}) \phi^{2} \right] + i\,
S_K[\bar{g}]
\right\} \, .  \label{6.11}
\eea
The conformal mode can now be integrated out. Because of the boundary
condition (\ref{6.5}), one must first make a shift
\beq
\phi (x)\,=\, \phi_{0}(x | \bar{g})\,+\, \varphi(x) \, , \label{6.12}
\eeq
where $\phi_{0}(x | \bar{g})$ is a solution of eqs.~(\ref{6.10}),
(\ref{6.5}),
and \beq
\varphi (x)\left|_{~\atop{{\cal I}^{-}, {\cal I}^{+}, I^{\circ}}}
=\, 0.   \right. \label{6.13}
\eeq
We have
\beq
\phi_{0} (x| \bar{g})\,=\, 1 \,+\, \frac{1}{6} \bar{G}\, \bar{{\cal R}} \, ,
\label{6.14} \eeq
where $\bar{G}$ is the Green function
\beq
\bar{G}\,=\, - \left( \bbox\,+\, \frac{1}{6}\, \bar{{\cal R}} \right) ^{-1} \,
,
\label{6.15} \eeq
whose boundary conditions, apart from the fact that the contraction $\bar{G}
\,\bar{{\cal R}}$ vanishes at infinity, are to be discussed below.

Equation (\ref{6.11}) takes the form
\bea
\lefteqn{
{\cal S}\,=\, \int [\di\bar{g}_{\mu \nu}]\, [\di \varphi ]\, (\det \bar{Q})
\delta (\chi (\bar{g}))\,  \exp \Biggl\{ i S_{conf}[\bar{g}]\, - \Biggr. }
\nnu \\
& &  -\frac{3i}{\kappa ^2} \int \di^{4}x \sqrt{- \bar{g}}\, \varphi
\left( \bbox
\,+\, \Biggl. \frac{1}{6}\, {\cal R}(\bar{g}) \right) \varphi \Biggr\}
\label{6.16} \eea
with
\beq
S_{conf}[\bar{g}]\,=\, - \frac{1}{2 \kappa^2}\int \di^{4}x \sqrt{- \bar{g}}
\left[ {\cal R} (\bar{g})\, +\, \frac{1}{6}\, {\cal R}(\bar{g})\, \bar{G}\,
{\cal R} (\bar{g})
\right]\,+\,  S_K[\bar{g}] \, . \label{6.17}
\eeq

The crucial point is the integration over the field $\varphi$. Its action has a
wrong sign, which is a consequence of the fact that, with Euclidean signature
of the metric, the Einstein-Hilbert action, even with the term $S_K[g]$
added, is
unbounded from below \cite{GHP}. It is the conformal mode that makes
it unbounded. The Euclidean functional integral over $\varphi$ diverges
exponentially and we must rotate the integration contour to
imaginary $\varphi$ \cite{GHP}. In the Lorentzian
context, the prescription to be used is different: one must take for
the propagator of the
field  $\varphi$, instead of the Feynman's Green function its complex
conjugate:
\beq
\left( \bbox\,+\, \frac{1}{6} \bar{{\cal R}} \right) \, \rightarrow\,
\left( \bbox \,+\,
\frac{1}{6} \bar{{\cal R}}\, - \, i \epsilon \right) \, . \label{6.18}
\eeq
One thus obtains
\beq
{\cal S}\,=\, \int [\di \bar{g}_{\mu \nu}]\, \delta (\chi(\bar{g}))\,
(\det \bar{Q})\, \left( \det (\bbox\,+\, \frac{1}{6} \bar{{\cal R}}\, - \, i
\epsilon) \right) ^{-1/2}\!\!\! \exp \Bigl( i S_{conf} [\bar{g}] \Bigr) \, .
\label{6.19} \eeq

The final step is the choice of the gauge-fixing function $ \chi (\bar{g})$
satisfying the condition $\det \bar{Q}\, \neq \,0$. Equation (\ref{6.19}) is
valid with any such $\chi (\bar{g})$, but let us choose
\beq
\chi(\bar{g})\,=\, {\cal R}(\bar{g})\, . \label{6.20}
\eeq
Then, for $\bar{Q}$ in eq.~(\ref{6.6}), one obtains
\beq
\bar{Q}(x,y)\,=\, 3 \left( \bbox - \frac{1}{3}\, \bar{{\cal R}} \right) \,
\delta(x,y) \label{6.21}\, .
 \eeq
Under the condition $\bar{{\cal R}} = 0$ implied by the delta-function
in  (\ref{6.19}), this guarantees that $\det \bar{Q}\, \neq\,0$. Furthermore,
under the  condition $\bar{{\cal R}} = 0$, one has from (\ref{6.17}):
\beq
 S_{conf}[\bar{g}]\left|_{~\atop \bar{R}=0} \,=\,S_K[{\bar{g}}]\right.\left
|_{~\atop \bar{R}=0}\right. \, , \label{6.22}
\eeq
and the functional integral (\ref{6.19}) takes the form
\beq
{\cal S}\,=\, \int [\di \bar{g}_{\mu \nu}]\, \delta ({\cal R}(\bar{g}))\, (\det
\bbox )^{1/2}\, \exp \Bigl( iS_K[\bar{g}] \Bigr)  \label{6.23}
\eeq
 with only the surface term in the action.

The action $S_{conf}[\bar{g}]$ in (\ref{6.17}) was first obtained in
\cite{canada} and next discussed in \cite{aaa,GHP}. It is invariant
under local conformal  transformations that become the identity at the
asymptotically flat infinity:

\beq
S_{conf}[\Omega^{2} \bar{g}]\,=\, S_{conf}[\bar{g}] \, . \label{6.24}
\eeq
Indeed, the action $S_{conf}[\bar{g}]$ is obtained from the action of
Einstein's
theory by making the substitution
\beq
g_{\mu \nu} (x)\,=\, \bar{g}_{\mu \nu}(x) \phi_{0}^{2} (x |\bar{g}) \, ,
\label{6.25}
\eeq
\beq
S_{conf}\, [\bar{g}]\,=\, S[g]\left|_{~\atop  g\,=\, \bar{g}
\phi_{0}^{2} (\bar{g})} \, , \label{6.26}
\right.
\eeq
\beq
S [g]\,=\, - \frac{1}{2 \kappa ^2} \int \di^{4} x\, \sqrt{- g}\,
{\cal R} (g)\,+\, S_K[g] \, . \label{SE}
\eeq
Since the function $\phi_{0}(x | \bar{g})$ in (\ref{6.14}) transforms as
\cite{canada,aaa}
\beq
\phi_{0}(x|\Omega^2 \bar{g})\,=\, \Omega^{-1}(x)\,\phi_{0}(x|\bar{g}) \, ,
\label{6.27}
 \eeq
the combination on the right-hand side of (\ref{6.25}) is conformal-invariant,
and  hence any functional of this combination is conformal invariant. The
action
$ S_{conf}[g]$ may be regarded as a conformal-invariant part of
Einstein's action. By rewriting eq.~(\ref{6.17}) in the form
\beq
S[g]\,=\, S_{conf}[g]\,+\, \frac{1}{12 \kappa ^2} \int \di^{4} x\,
\sqrt{- g}\: {\cal R} \,G\,{\cal R} \, , \label{6.28}
\eeq
one makes it obvious that Einstein's
action is the broken action $S_{conf}$ with the gauge-fixing function
(\ref{6.20}) introduced in the action  quadratically. Since the action
$S_{conf}
[g]$ does not contain the harmful conformal mode, one may expect that, with
Euclidean signature of the metric, this action is already positive-definite.
This is indeed the case.
In fact, by using the conformal invariance, the action $S_{conf}[g]$
can be brought to the gauge ${\cal R} = 0$ where
\beq
S_{conf} [g]\,=\, S_K[g] \left|_{~\atop R\,=\,0\,} \label{6.29}
\right. \, ,
\eeq
and it has been proved \cite{SY} that
\beq
-S_K[g] \left|_{~\atop R=0}\,\geq\, 0, ~~ \mbox{sign}\: g\,=\, (+ + + +) \, .
 \right. \label{6.30}
\eeq

In the derivation above, the final result in eq.(\ref{6.23}) appears to be
obtained by
identically transforming the original functional integral. However, even
apart from the problem with the propagation of the conformal mode, one
reserve remains:
the bar on the integration variables $\bar {g}_{\mu\nu}$ in eqs.~(\ref{6.19})
and  (\ref{6.23}) is kept because
the asymptotic fields for $g_{\mu \nu}$ and $\bar{g}_{\mu \nu}$ need not
be one and the same. Equation (\ref{6.19}) looks like the result of the
quantization
of  the invariant action $S_{conf}$ obtained by the usual rules of gauge theory
with $\delta (\chi)$ the gauge-breaking term and det$Q$ the ghost term. The
only apparent difference from the usual rules is that, because the action
$S_{conf}$ is non-local, there appears an additional invariant measure
\beq
\left( \det \left( \bx \,+\, \frac{1}{6}\, {\cal R}\, - \, i \epsilon \right)
\right)^{-1/2} \, . \label{6.31} \eeq
 However, gravity theory is not conformal-invariant, and the choice of the
conformal gauge is relevant for the physical metric. There is no
other place in the integral (\ref{6.19}) except the asymptotic condition where
the conformal invariance can be broken. On the other hand, there is one
and only one choice of the conformal gauge with which the metric measures
distance. For pure gravity or gravity coupled to conformal-invariant matter
fields, this choice is ${\cal R}\,=\,0$. At the classical level, the
variational
equations of the action $S_{conf}[g]$, supplemented with the equation
${\cal R} = 0$,
are equivalent to the Einstein equation. It is, therefore, plausible that, for
$\bar{g}_{\mu\nu}$ in (\ref{6.23}), the original asymptotic conditions apply
and, generally, $\bar{g}_{\mu \nu}$ in (\ref{6.23}) can be identified with the
operator  of the physical metric. The functional integral (\ref{6.23}), or
its
generalization to the presence of massless matter, could then be used
directly for calculating the expectation values of the
gravitational-field  observables and the scattering matrix.

The relevance of the result (\ref{6.23}) is that, after integrating out the
conformal mode, the action in the functional integral becomes a pure surface
term. Therefore, this action can be written again in terms of Bondi masses,
those pertaining, this time, to
 the virtual gravitational fields
that are summed over in the functional integral.
By extending the approach advocated in this paper
 one can try to use the above result to account
for the contributions to the scattering amplitude from non-perturbative virtual
  field configurations.

\newpage

{\Large \bf Acknowledgments}
\bigskip

 We are grateful to  D. Amati, M. Ciafaloni and N.~Sanchez for
interesting discussions. R.P. and   G.A.V. acknowledge
the CERN Theory Division for the kind
hospitality extended to them. G.A.V. also thanks
the Physics Department at the University of Naples, where part of this
work was done.

R.P. has been partially  supported in the framework
of the E.C. Research Program ``Gauge theories, applied Supersymmetry
and Quantum Gravity'' with a financial contribution under contract
SC1-CT92-0789. G.A.V. has been supported by the Russian Science
Foundation under grant 93-02-15594.

\newpage
{\Large \bf Appendix}
\appendix

\section{Calculation of the boundary term}
\setcounter{equation}{0}

\qquad
  Let us introduce on each null surface, where $u$ is
 constant, two coordinates
$\phi^{a},\, a=1,2$,  which label the light rays, and a third coordinate
$r$, which is a parameter  along the rays. The coordinates $\phi^{a}$ take
values on a 2-sphere and can be chosen so as to ensure  orthogonality,
that is $(\nabla u, \, \nabla \phi^{a}) \equiv 0$. The parameter $r$ can be
chosen to be the luminosity distance along the rays by considering the
induced metric~\fnote{\dag}{The induced metric on $u\,=\, \mbox{const}$ is
two-dimensional because, along the rays, the interval is zero.} on
$u=\mbox{const}$\,: \beq
\di s^{2} \left|_{u=const}\,=\, g_{ab}\, \di \phi^{a} \di \phi^{b}\, ,
\hspace{2cm}  \,(a,b\,=\, 1,2)\, ,\right.
\eeq
where
\beq
g_{ab}\,=\, g_{ab}(\phi, r, u)\left|_{u=const}\,,\right.
\eeq
and requiring that the area of the two-dimensional section
$u=\mbox{const},\,r=\mbox{const}\,$ be
\beq
 \int_{\!\! ~\atop{2-sphere}}\di ^{2} \phi \sqrt{
\mbox{det} \, g_{ab}}\,=\, 4 \pi
 r^{2} \, .
\label{A.2}
\eeq

The functions $u,\,r,\,\phi^{a}$ can serve as local coordinates. Four
components of the metric in this coordinate frame are already known. Three
of them are
\bea
g^{uu} & = & (\nabla u)^{2}\, \equiv 0 \\
g^{u \phi^{1}} & = & (\nabla u, \, \nabla \phi^{1}) \equiv 0, \\
g^{u \phi^{2}} & = & (\nabla u, \, \nabla \phi^{2}) \equiv 0,
\eea
and the fourth is fixed by the constraint (\ref{A.2}). The remaining six
components of the metric characterize the gravitational field. We introduce
six unknown functions $\Psi, \, W,\,  \gamma, \, \delta$ and
$ U^{a} \, (a\, =
\, 1,2)$, so that
\bea
g^{ur} & = & (\nabla u,\, \nabla r)\,=\, \frac{1}{\Psi}, \\
& &  \nnu \\
g^{rr} & = & (\nabla r)^{2}\,=\, \frac{1}{\Psi^{2}} W,  \label{A.7}\\
& &   \nnu \\
g^{r \phi^{a}} & = & (\nabla r, \, \nabla \phi^{a})\,=\, \frac{1}{\Psi}
\, U^{a} \, ~~~~~~~~( a\,=\, 1,2) \, ,
\eea
where
\beq
\phi^{1}  =  \theta\, , \: \phi^{2}\,=\, \varphi \qquad
0\, \leq\, \theta \, \leq \, \pi,~~ 0\, \leq \varphi \, \leq\, 2 \pi \, ,
\eeq
and
\bea
(\nabla r, \, \nabla \theta) & = & \frac{1}{\Psi} U^{\theta}, \hspace{1cm}
(\nabla r,
\nabla \varphi)\,=\, \frac{1}{\Psi} U^{\varphi} \, .
\eea
Next
\beq
g^{\phi^{a}\phi^{b}}  =  (\nabla \phi^{a},\, \nabla \phi^{b})\,=\,
g^{ab} =\, (g_{ab})^{-1} \, ,
\eeq
where
\bea
g_{ab} \, \di \phi^{a} \di \phi^{b} & = & r^{2}\left[ \frac{1}{2} \,(e^{2
\gamma}\,+\, e^{2 \delta}) \di \theta^{2}\,+\,
 (e^{\gamma - \delta}- \, e^{\delta - \gamma})\sin \theta\,\di \,
\theta\,\di \, \varphi \right. \nnu \\
&  & + \left. \frac{1}{2}(e^{-2 \gamma}+ \, e^{-2 \delta}) \sin^{2} \theta
\di \, \varphi^{2} \right]\, , \label{A.11}
\eea
and
\beq
\det g_{ab}\,=\, r^{4}\sin^{2}\theta \, ,
\eeq
in agreement with (\ref{A.2}). Then the metric takes the form
\beq
\di s^{2} \, = \, -\, W\, \di u^{2}\,+\, 2 \Psi \di u \,\di r\, +\,
g_{ab} (\di \phi^{a}\,-\, U^{a} \di u)\, (\di \phi^{b} - U^{b} \di u)
\label{A.13}
\eeq
with $g_{ab}$ given by  (\ref{A.11}), and its determinant is
\beq
g\,=\, \det g_{\mu \nu}\,=\, -\, \Psi^{2} r^{4} \sin^{2}\theta \, .
\label{A.14}
 \eeq

As seen from (\ref{A.14}), the chart $u,\, r,\, \phi^{a}$ breaks
down when $\Psi$ becomes zero or infinity. The reason for that can be
understood by considering the null geodesic $u\,=\mbox{const}, \, \phi^{a}\,
=\,\mbox{const}$. Let $\lambda$ be an affine parameter along this geodesic,
growing towards the future. From the geodesic equation one finds
\beq
\frac{\di r}{\di
\lambda}\left|_{~\atop{{u=const}\atop{\phi=const}}} \propto
 (\nabla u, \, \nabla r)=
\frac{1}{\Psi}\right. \, ,   \eeq
from which it follows that, at the point where $\Psi$ becomes  zero or
infinity, $r$ stops being a monotonic parameter along the light ray passing
through this point. At this point the light ray hits the
apparent horizon. If, during the
history of an  outgoing light ray, $\di r/ \di \lambda$ changes its sign an odd
number of times  or if $\di r/\di \lambda \rightarrow 0$ as $\lambda
\rightarrow \infty$, this  light ray will never reach the asymptotic domain
of infinite luminosity distance, where the metric becomes flat. By
definition, this light ray will not come to ${\cal I}^{+}$.

By considering only a portion of the full congruence of null geodesics that
come to ${\cal I}^{+}$, we automatically guarantee that the chart $u, \, r,\,
\phi^{a}$ covers the asymptotic domain near ${\cal I}^{+}$.
Equation (\ref{A.13}) gives the general form of an asymptotically flat metric
near  ${\cal I}^{+}$, and ${\cal I}^{+}$ is reached at the limit $r
\rightarrow \infty$ with fixed $u$ and $\phi^{a}$. The remaining
arbitrariness in the choice of the variable $u$ is fixed by the
normalization condition
\beq
(\nabla u, \, \nabla r)\left|_{{\cal I}^{+}}\,=\, - 1\right. \, .
\label{A.16} \eeq
Then
\beq
(\nabla u, \, \nabla r)\,=\, \frac{1}{\Psi} < 0 \,
\eeq
everywhere in the domain covered by the chart $u,\,r,\,\phi^{a}$.
The $u$ and $\phi^a$
 act as
 coordinates on  ${\cal I}^{+}$, which is topologically a product of the
time axis by the 2-sphere. Because of (\ref{A.16}), the retarded time $u$ grows
along ${\cal I}^{+}$ towards the future and coincides with the proper time of
an
observer at rest at large and constant $r$. The   metric at
 ${\cal I}^{+}$  behaves as follows $(r \rightarrow
\infty  )$
\beq
W \left|_{~\atop{{\cal I}^{+}}}\,=\,1\, -\, {2\, G {\cal M}_{+}(u, \varphi,
\theta)\over{r}}\,
+\, {\cal O} \left( {{1}\over{r^{2}}}\right) \, , \right.
\label{A.18}
\eeq
\beq
\frac{\gamma + \delta}{2} \left|{~\atop{{\cal I}^{+}}}
\, = \, {{C_{1}^{+}
(u,\varphi, \theta)}\over {r}}\, +\, {\cal O} \left( {{1}\over{r^{2}}}
\right) \, ,
\right. \label{A.19}
\eeq
\beq
\frac{\gamma - \delta}{2}\left|_{~\atop{{\cal I}^{+}}} \, =\,
{{C_{2}^{+}(u, \varphi,
\theta)}\over{r}}\,+\, {\cal O} \left( {{1}\over{r^{2}}}
\right) \, , \right. \label{A.20}
\eeq
\beq
U^{\theta}\left|_{~\atop{{\cal I}^{+}}} \,=\, {2 \,{\cal
N}_{+}^{\theta}(u,\varphi,
\theta)\over {r^{2}}}\,+\,
{\cal O} \left( {{1}\over{r^{3}}}
\right) \, , \right.
\label{A.21}
\eeq
\beq
U^{\varphi}\left|_{~\atop{{\cal I}^{+}}} \,=\, {{2 \,{\cal N}_{+}^{\varphi}
(u, \, \varphi,\, \theta)}\over{r^{2}}}\,+\, {\cal O} \left( {{1}\over{r^{3}}}
\right) \, ,
 \right. \label{A.22}
\eeq
\beq
\Psi \left|_{~\atop{{\cal I}^{+}}} \,=\, - 1\,+\,
{\cal O} \left( {{1}\over{r^{3}}} \right) \, ,
\right. \label{A.23}
\eeq
where ${\cal M}_{+}, C_{1}^{+}, C_{2}^{+}, {\cal N}_{+}^{\theta},
{\cal N}_{+}^{\varphi}$ are finite functions of $u, \, \varphi, \theta$.

According to (\ref{A.7}),
\beq
(\nabla r)^{2} \left|_{~\atop{{\cal I}^{+}}} \,=\,   1\,-\,
{2\, G{\cal M_{+}}
(u, \varphi, \theta)\over{r}} \,+\,{\cal O}
\left( {{1}\over{r^{2}}} \right) \, ,\right.
\label{A.24} \eeq
and ${\cal M}_{+}$ averaged over the two-sphere gives
\beq
M_{+} (u)\,=\, \frac{1}{4 \pi} \int_{0}^{2 \pi} \di \varphi \int_{0}^{\pi}
\di \theta \sin \theta \, {\cal M_{+}}(u, \varphi, \theta) \, , \label{A.25}
\eeq
which is the Bondi mass at ${\cal I}^{+}$, that is, the energy that
remains in the system by the instant $u$ of retarded time. Its limit at
$u \rightarrow - \infty$,
\beq
M_{+} (-\, \infty)\,=\, M_{0} \,
\eeq
is the $ADM$ mass which is the full conserved energy stored in space-time
and measured at spatial infinity $I^{\circ}$ . The difference
\beq
M_{0}\,-\, M_{+} (u)\,=\, \int_{- \infty}^{u} \di u \, \left( - \frac{\di
M_{+}}
{\di u} \right) \,
\eeq
is the energy radiated away through ${\cal I}^{+}$ by the instant $u$ of
retarded time. This is a sum of the energy carried away by the gravitational
waves and the energy radiated by massless sources. The energy conservation law
following from the Einstein equations in (\ref{2.5}) is of the form
\bea
- \frac{\di}{\di u} M_{+}(u) & = & \frac{1}{4 \pi} \int_{0}^{2 \pi}\di
\varphi  \int_{0}^{\pi} \di \theta \sin \theta
\left[ \left( \frac{\partial}{\partial u} C_{1}^{+} \right) ^{2} \, +\,
\left( \frac{\partial}{\partial u} C_{2}^{+}\right) ^{2}\right] \nnu \\
& &
+ \left. \int_{0}^{2 \pi} \di \varphi \int_{0}^{\pi} \di \theta \sin \theta
\left[ \frac{1}{4}\, r^{2}\, {\cal T}_{\mu \nu} \nabla^{\mu} v \nabla^{\nu}v
\right] \right|_{{\cal I}^{+}}
 \label{A.28}
\eea
where $v\,=\, \mbox{const}$ is the null congruence parallel to ${\cal
I}^{+}$: \beq
(\nabla v)^{2}\, \equiv \,0\,, ~~(\nabla v, \, \nabla u) \mid_{\cal I^{+}}
\,=\, -\, 2 \, . \label{A.29}
\eeq
The derivatives $ \partial C_{1}^{+}/ \partial u$ and $ \partial C_{2}^{+}
/\partial u$ of the functions appearing in (\ref{A.19}), (\ref{A.20}) and
(\ref{A.28}) are  the Bondi-Sachs news functions. They determine the
manifestly positive energy  flux of the gravitational radiation. They are
also the data (final data  in the case of ${\cal I}^{+}$, initial data in
the case of ${\cal I}^{-}$) for the two degrees of freedom of the
gravitational field counted for a three-dimensional point. The
three-dimensional
points for which the degrees of freedom of the  asymptotically flat
gravitational field are counted are points of ${\cal I}^ {+}$ or ${\cal
I}^{-}$. Finally, the last term in (\ref{A.28}) is the energy flux of the
source. It will be non-vanishing at ${\cal I}^{+}$ only if the source
contains a massless component.

The metric near ${\cal I}^{-}$ is of a form similar to (\ref{A.13}) but, in
this case, one considers a converging null congruence
\beq
(\nabla v)^{2} \equiv 0\, , ~~ (\nabla v, \, \nabla r)\mid_{\cal I^{-}}~
=1 \, , \label{A.30}
\eeq
and the limit of infinite luminosity distance $r$ is reached by going along
the light rays towards the past. Equations similar to
(\ref{A.18})--(\ref{A.20}) and (\ref{A.24})--(\ref{A.25}) define the Bondi
mass $M_{-}(v)$ and news functions $\partial  C_{1}^{-}/ \partial v , ~
\partial C_{2}^{-}/ \partial v$ at ${\cal I}^{-}$. The Bondi mass $M_{-}
(v)$ is the energy brought to the system by the instant $v$ of advanced
time, and its limit at $v \rightarrow + \infty$ is the full  $ADM$ mass:
\beq
M_{-} (+ \infty) \,=\, M_{0} \, .
\eeq
The difference $M_{0} \, - \, M_{-} (v)$ is the energy brought to the system by
the incoming radiation after the instant $v$ of advanced time. Again, this
is a sum of the energies of the incoming gravitational radiation and of an
incoming flux of massless sources. The energy conservation law at
${\cal I}^{-}$ is obtained from (\ref{A.28}) by changing the notation and
the  sign of the left-hand side.

Of interest are also the limits $M_{-} (- \infty)$ and $M_{+} (+\,\infty)$.
Since we have
\beq
M_{0}\,=\, M_{-}(- \infty)\,+\, \int_{- \infty}^{\infty} \di v \:
\frac{\di M_{-}} {\di v} \, ,
\eeq
\beq
M_{0}\,=\, M_{+} (+ \infty)\,+\, \int_{- \infty}^{\infty} \di u \:
\left( - \frac{\di
M_{+}} {\di u} \right) \,
\eeq
and the integrals on the right-hand sides are, respectively, the energy brought
by radiation and the energy carried away by radiation during the whole history,
we conclude that $M_{-}(- \infty)$ is the energy brought to the system by
massive (time-like) sources coming from $I^{-}$. Similarly $M_{+} (+ \infty)$
is the remainder that goes to $I^{+}$ with time-like sources after the total
emission of radiation.

 By assuming that ${\cal T}^{\mu \nu}$ satisfies the dominant energy condition,
it  has been proved \cite{Wald}
that not only the radiation fluxes, both gravitational and matter, are
non-negative, but also that the $ADM$ mass and Bondi masses are always
non-negative, which means, in particular, that the asymptotically flat system
cannot radiate more energy than initially stored.

The functions ${\cal N}_{+}^{\theta}$  and ${\cal N}_{+}^{\varphi}$ appearing
in
eqs. (\ref{A.21}) and (\ref{A.22}) and the functions ${\cal N}_{-}^{\theta}$
and  ${\cal N}_{-}^{\varphi}$ similarly defined at ${\cal I}^{-}$, are
associated  with radiated angular momentum.

 To calculate the surface integral (\ref{2.13}) over the boundary
\beq
{\cal I}^{-} \cup {\cal I}^{+}\,:\, \tau(x)\,=\, 0 \,
\eeq
one may first go over to the Penrose space where ${\cal I}^{-}$ and ${\cal I}^
{+}$ really exist as three-dimensional null surfaces, but one may also do this
calculation directly in the physical space-time by considering the null
surfaces close to infinity and, next, going over to the limit of
${\cal I}^{-}$ or ${\cal I}^{+}$ respectively. The contribution of
${\cal I}^{+}$ is then of the form
\beq
S_K\mid_{\cal I^{+}} \, = \, \lim_{c \rightarrow 0} \frac{1}{16 \pi G}
\int \di ^{4} x \sqrt{- \tilde{g}}\, \delta \left(- \frac{1}{v(x)}\,+\,c
\right)
 A^{\mu} \tilde{\nabla}_{\mu} \left( - \frac{1}{v(x)}\,+\, c \right)\, ,
\label{A.35}
\eeq
with $v(x)$ defined in eq. (\ref{A.29}), and
\beq
A^{\mu}\,=\, \tilde{g}^{\alpha \beta} \tilde{\nabla}^{\mu} h_{\alpha \beta}\,
-\, \tilde{g}^{\mu  \alpha} \tilde{\nabla}^{\beta} h_{\alpha \beta} \, .
\eeq

We shall  carry out the calculation of (\ref{A.35}) by using the Bondi-Sachs
frame,
where the metric has the form (\ref{A.13}). Accordingly, the flat metric
$\tilde{g}_{\mu \nu}$ has the form
\beq
\tilde{g}_{\mu \nu} \di x^{\mu} \di x^{\nu} \, = \, - \di u^{2}\, -
 \, 2 \di u \di
\,r\, + \, r^{2} \left(\di \theta^{2}\,+\, \sin^{2} \theta \di \varphi^{2}
\right), \label{A.36}
\eeq
and the components of $h_{\alpha \beta}$ that do not vanish identically can be
read off from (\ref{A.18})--(\ref{A.23}):
\bea
h_{uu} & = & \frac{2\, G {\cal M}_{+}}{r}\,+\,
{\cal O} \left( {{1}\over{r^{2}}} \right) \, ,
\label{A.38}\\
 h_{ur} & = & h_{ru}\,=\, {\cal O}
\left( {{1}\over{r^{3}}} \right) \, ,\\
h_{u \theta} & = & h_{\theta u}\,=\, - 2 \,{\cal N}_{+}^{\theta}\,+\,
{\cal O} \left( {{1}\over{r}} \right) \, , \\
h_{u \varphi} & = & h_{\varphi u} \,=\, - 2 \sin^{2} \theta
\, {\cal N}_{+}^{\varphi}\,
+\,
{\cal O} \left( {{1}\over{r}}\right) \, , \\
h_{\theta \theta} & = & r^{2} \left(\frac{2\, C_{1}^{+}}{r}\,
+\,
{\cal O} \left( {{1}\over{r^{2}}}\right) \right) \, , \\
 h_{\varphi \varphi} & = & r^{2} \sin^{2} \theta \left(- \frac{2\,
C_{1}^{+}}{r}\, +\,
{\cal O} \left( {{1}\over{r^{2}}}\right) \right) \, ,\\
h_{\theta \varphi} & = & h_{\varphi \theta} \,=\, r^{2} \sin \theta
\left(  \frac{2\, C_{2}^{+}}{r}\,+\,{\cal O}
\left( {{1}\over{r^{2}}}\right) \right)
\, .  \label{A.44} \eea

In the asymptotic domain near ${\cal I}^{+}$, eqs.~(\ref{A.29}) are solved
by \beq
v(x)\,=\, 2 r \,+\, u,~~~r \rightarrow \infty~~\mbox{at fixed}\,\, u.
\eeq
This makes it possible to carry out the integral over $r$ in (\ref{A.35}) with
the
aid of the delta-function, and the result is the limit $r \rightarrow
\infty$ of the remaining three-dimensional integral:
\beq
S_K \left|_{~\atop\cal I^{+}} \, =\, \lim_{r \rightarrow \infty} \frac{1}
{16 \pi G}
\int_{0}^{2 \pi} \di \varphi \int_{0}^{\pi} \di \theta \sin \theta \int_{-
\infty} ^{\infty} \di u \, r^{2} A^{\mu} \tilde{\nabla}_{\mu} \left(
r +
\frac{1}{2}\, u \right) \, . \right. \label{A.46}
\eeq
For the integrand in (\ref{A.46}) we have
\beq
A^{\mu} \tilde{\nabla}_{\mu} \left(r + \frac{1}{2}\, u
\right) \,= \, \left( \frac{1}{2}
\frac{\partial}{\partial r} - \frac{\partial}{\partial u}
\right) (\tilde{g}^{\alpha
\beta} h_{\alpha \beta}) - \left( \frac{1}{2}\, \delta_{r}^{\alpha} -
\delta_{u}^{\alpha}\right) (\tilde{\nabla}^ {\beta} h_{\alpha \beta}) \, .
\label{A.47} \eeq

By calculating the covariant derivatives in the metric (\ref{A.36}), one
obtains
\bea
\delta_{r}^{\alpha} \left(\tilde{\nabla}^{\beta} h_{\alpha \beta}\right)
& = & -
\partial_{r} h_{u r} \,-\, \frac{2}{r}\, h_{u r}\,-  \frac{1}{r^{3}}\,
h_{\theta
\theta}\,-\, \frac{1}{r^{3} \sin^{2}\, \theta} h_{\varphi \varphi} \, , \\
\delta_{u}^{\alpha} (\tilde{\nabla}^{\beta} h_{\alpha \beta})
& = & -
\partial_{r} h_{uu} - \frac{2}{r}\, h_{uu} \,+\, \partial_{r}\, h_{ur} -
\partial_{u} h_{ur}\,+\, \frac{2}{r}\, h_{ur}\,+\, \frac{1}{r^{2}}\,
\partial_{\theta} h_{u \theta}
\nnu \\
& &
 + \frac{1}{r^{2}} \,\frac{\cos \theta}{\sin \theta}\, h_{u
\theta}\,+\, \frac{1}{ r^{2} \sin^{2} \theta} \,\partial_{\varphi} h_{u
\varphi}
\, ,
\eea
and
\beq
\tilde{g}^{\alpha \beta} h_{\alpha \beta}\,=\, - 2\, h_{ur}\,+\,
\frac{1}{r^{2}}
\, h_{\theta \theta} +\, \frac{1}{r^{2} \sin^{2} \theta}\, h_{\varphi \varphi}
\,
. \eeq
It is now easy to insert the asymptotic behaviors (\ref{A.38})--(\ref{A.44})
to get, for (\ref{A.47}):
\beq
 \lim_{r \rightarrow \infty} r^{2} A^{\mu} \tilde{\nabla}_{\mu}
\left( r +
\frac{1}{2} u \right)
 = - 2 \,G{\cal M}_{+} - 2\, \partial_{\varphi} {\cal N}_{+}^{\varphi} - 2
\,\frac{1}
{\sin \theta} \,\partial_{\theta} \left( \sin \theta {\cal N}_{+}^{\theta}
\right) \, .
\eeq
Upon integration over the  sphere in (\ref{A.46}), the contribution of
${\cal N}_{+} ^{\varphi}, {\cal N}_{+}^{\theta}$ vanishes, and we obtain
\beq
S_K\left|_{~  \atop{\cal I^+}} \, =\, - \frac{1}{8 \pi} \int_{0}^{2 \pi}
\di  \varphi
\int_{0}^{\pi} \di \theta \sin \theta \int_{- \infty}^{\infty} \di u\, {\cal
M}_{+} (u, \varphi, \theta)  \, . \right.
\eeq
A similar result, but with the opposite sign, is obtained at ${\cal I}^{-}$.
The final expression for $S_K [g]$ can thus be presented in the form
\bea
S_K & = & - \frac{1}{8 \pi} \left(\int_{\cal I^{+}} {\cal M}_{+} - \int_{\cal
I^{-}}\, {\cal M}_{-} \right) \nnu \\
&  & \nnu\\
& = &- \frac{1}{2} \left( \int_{- \infty}^{\infty} \di u\, M_{+} (u) -
\int_{- \infty}^{\infty} \di v\, M_{-}(v) \right)\, ,
\eea
where $M_{+}(u)$ and $M_{-}(v)$ are the Bondi masses at ${\cal I}^{+}$ and
${\cal I}^{-}$, and the retarded and advanced time are normalized by the
conditions (\ref{A.16}) and (\ref{A.30}).

\newpage
\renewcommand{\baselinestretch}{1}

}
\newpage
{\bf Figure Captions}

\bigskip

{\bf Fig. 1:} Tree diagrams (after cutting the scalar
propagators) which are resummed by our method in the case of elastic
scattering. The blob represents all (including disconnected) trees.

\bigskip

{\bf Fig. 2:} Same as in fig. 1 for the case of inelastic processes.

\end{document}